\documentclass[preprint,aps,showpacs,nofootinbib]{revtex4}

\usepackage{epsfig,amssymb,amsmath}

\def\bea{\begin{eqnarray}}
\def\eea{\end{eqnarray}}
\def\bec{\begin{center}}
\def\ec{\end{center}}

\def\beq{\begin{equation}}
\def\eeq{\end{equation}}

\begin{document}
\draft \tighten \preprint{KAIST-TH 2006/12} \preprint{KUNS-2056}
\preprint{KYUSHU-HET-102}
\title{\large \bf TeV Scale Mirage Mediation and Natural Little SUSY Hierarchy}
\author{
Kiwoon Choi\footnote{kchoi@hep.kaist.ac.kr}$^1$,
Kwang Sik Jeong\footnote{ksjeong@hep.kaist.ac.kr}$^1$,
Tatsuo Kobayashi\footnote{kobayash@gauge.scphys.kyoto-u.ac.jp}$^2$ and
Ken-ichi Okumura\footnote{okumura@higgs.phys.kyushu-u.ac.jp}$^3$
}
\address{
$^1$Department of Physics, Korea Advanced Institute of Science
and Technology, Daejeon 305-701, Korea \\
$^2$Department of Physics, Kyoto University,
Kyoto 606-8502, Japan \\
$^3$Department of Physics, Kyushu University, Fukuoka 812-8581,
Japan}

\begin{abstract}
TeV scale mirage mediation has been proposed as a supersymmetry
breaking scheme reducing the fine tuning for  electroweak symmetry breaking in
the minimal supersymmetric extension of the standard model.
We discuss a moduli stabilization set-up for TeV scale mirage mediation which
allows an extra-dimensional interpretation for the origin of supersymmetry
breaking and naturally gives an weak-scale size of the Higgs $B$-parameter.
The set-up utilizes the holomorphic gauge kinetic functions depending on both
the heavy dilaton and the light volume modulus whose axion partners are assumed
to be periodic fields.
We also examine the low energy phenomenology of  TeV scale mirage mediation,
particularly the constraints from electroweak symmetry breaking and FCNC
processes.
\end{abstract}
\maketitle

\section{introduction}

Low-energy supersymmetry (SUSY) is one of the primary candidates for physics
beyond the standard model (SM) above the weak scale \cite{Nilles:1983ge}.
One strong motivation for supersymmetric extension of the SM is to solve the
hierarchy problem between the weak scale and GUT/Planck scale.
In particular, the minimal supersymmetric standard model (MSSM) is quite
interesting from the viewpoint of its minimality as well as the realization
of gauge coupling unification at $M_{GUT} \sim 2 \times 10^{16}$ GeV.

In supersymmetric standard model, the lightest Higgs boson $h^0$ is predicted
to have a  light mass. Including the one-loop correction while ignoring the
effect of stop mixing \cite{Haber:1990aw}, $m_{h^0}$ in the MSSM is given by
\bea
m_{h^0}^2\simeq M_Z^2\cos^2
2\beta+\frac{3y_t^2m_{\tilde{t}}^2}{4\pi^2}
\ln(m_{\tilde{t}}^2/m_t^2),
\eea
where $M_Z$ is the $Z$-boson mass,
$\tan\beta=\langle H^0_u\rangle/\langle H^0_d\rangle\gtrsim 3$, $y_t$
is the top quark Yukawa coupling, and $m_{\tilde{t}}$ is the stop mass.
Thus the current experimental bound for the SM like Higgs $m_{h^0}>114$ GeV can
be satisfied within the MSSM,
but it implies a rather heavy stop mass, e.g. $m_{\tilde{t}}\gtrsim 600$ GeV.
In supersymmetric models, $m_{\tilde{t}}$ is tightly linked to the up-type
Higgs soft mass $m_{H_u}$ through the renormalization group (RG) evolution
induced by the large value of $y_t$:
\bea
\label{rgmixing}
\delta m_{H_u}^2 \sim - \frac{3y_t^2m_{\tilde
t}^2}{4\pi^2}\ln(\Lambda/m_{\tilde{t}}),
\eea
where $\Lambda$ is the (effective) messenger scale of SUSY breaking which is
expected to be close to the GUT/Planck scale in generic high scale mediation
models.
Unless cancelled by other effects, this RG evolution implies that
$|m_{H_u}^2|\sim m_{\tilde{t}}^2$ at the weak scale.
On the other hand, the electroweak symmetry breaking (EWSB) conditions in the
MSSM give rise to
\bea
\label{ewsb}
\frac{M_Z^2}{2} \simeq
-\mu^2(M_Z)-m_{H_u}^2(M_Z)+\frac{m_{H_d}^2(M_Z)}{\tan^2\beta},
\eea
where $\mu$ is the Higgsino mass and $m_{H_d}$ is the down-type Higgs soft
mass.
This EWSB condition requires a fine tuning of parameters with an accuracy of
${\cal O}(1)\, \%$ if $m_{H_u}$ is heavier than 600 GeV as suggested by the
lower bound of $m_{h^0}$ and the RG evolution of $m_{H_u}^2$.
This is the so-called little SUSY hierarchy problem \cite{Barbieri:1987fn}.

During the last years, several types of scenarios solving the little SUSY
hierarchy problem have been proposed \cite{casas}--\cite{kaplan}.
Many of them extend the MSSM to increase $m_{h^0}$ while keeping the
superparticle masses as light as possible.
Alternative possibility is to have a particular pattern of SUSY breaking soft
terms within the MSSM \cite{Choi:2005uz,hdkim},
satisfying the EWSB condition (\ref{ewsb}) without fine tuning.
A particularly interesting proposal along this direction is the TeV scale mirage
mediation of SUSY breaking \cite{Choi:2005hd,Kitano:2005wc} which gives a little
hierarchy between $m_{H_u}$ and $m_{\tilde{t}}$ in a natural
manner\footnote{The schemes proposed in \cite{hdkim1,nomura1} also give a
qualitatively similar pattern of soft terms.}:
\bea
\label{little}
|\,m_{H_u}^2(M_Z)| \,\sim\,
\frac{m_{\tilde{t}}^2(M_Z)}{8\pi^2}.
\eea

In mirage mediation \cite{choi1}, anomaly-mediated SUSY breaking \cite{anomaly}
and modulus-mediated SUSY breaking \cite{modulus} are dynamically arranged to
cancel the RG evolution of soft parameters \cite{Choi:2005uz}.
Such pattern of SUSY breaking is a natural outcome of KKLT-type moduli
stabilization \cite{Kachru:2003aw} in which the modulus $F$-component is
suppressed compared to the gravitino mass $m_{3/2}$ by the factor
$1/\ln(M_{Pl}/m_{3/2})$ \cite{choi1}.
The typical size of superparticle masses in this scheme is given by
\bea
m_{\rm SUSY} \sim \frac{m_{3/2}}{8\pi^2},
\eea
while the detailed pattern is determined by the anomaly to modulus mediation
ratio.
Under certain assumption on the discrete parameters of underlying theory,
the effective RG evolution of soft parameters in mirage mediation is determined
by a "mirage messenger scale"
\bea
\Lambda\sim M_{\rm mir} \equiv
\frac{M_{GUT}}{(M_{Pl}/m_{3/2})^{\alpha/2}},
\eea
where
$\alpha={\cal O}(1)$ parameterizes the anomaly to modulus mediation ratio
\cite{Choi:2005uz}.
Having $\alpha=2$ leads to $M_{\rm mir}\sim 1$ TeV minimizing the effective RG
evolution of $m_{H_u}^2$, thereby allows the little hierarchy (\ref{little})
realized without fine tuning.
The TeV scale mirage mediation solving the little hierarchy problem can give two
different mass patterns at the weak scale suggested by the EWSB condition
(\ref{ewsb}):
\bea
\label{masspattern}
&\mbox{(I)}& \quad
\mu\,\sim\, m_{H_{u,d}} \sim M_Z,\quad m_{\tilde{t}}\sim
\sqrt{8\pi^2}M_Z, \quad B\sim M_Z/\tan\beta,
\nonumber \\
&\mbox{(II)}& \quad
\mu\,\sim\, m_{H_u}\sim M_Z,\quad
m_{H_d}\sim m_{\tilde t}\sim \sqrt{8\pi^2}
M_Z, \quad B\sim 8\pi^2M_Z/\tan\beta,
\eea
where we have used another EWSB condition $\mu B\simeq
(m_{H_d}^2+m_{H_u}^2+2\mu^2)/\tan\beta$ for the estimate of the Higgs mass
parameter $B$.
In Ref.~\cite{Choi:2005hd}, it has been shown that both mass patterns can be
obtained in a certain class of (string-motivated) effective supergravity
(SUGRA) model with SUSY-breaking uplifting potential.
The same model giving the mass pattern (I) has been discussed also in
\cite{Kitano:2005wc}, followed by a phenomenological study including the degree
of fine-tuning, dark matter detection and collider signals \cite{Kitano:2005ew}.

Recently, it has been pointed out that the uplifting potential which has been
assumed in \cite{Choi:2005hd,Kitano:2005wc} to get $\alpha=2$ is difficult to
have an extra-dimensional interpretation \cite{pierce}.
This would cast a doubt on the naturalness of the whole set-up.
Indeed, if the uplifting potential originates from a SUSY-breaking brane
stabilized at the IR end of warped throat as in the KKLT moduli stabilization
scenario, the minimal set-up discussed in \cite{choi1,Choi:2005uz} gives
$\alpha=1$, and thus an intermediate scale value of $M_{\rm mir}$.
In this paper, we propose an alternative scheme giving $M_{\rm mir}\sim 1$ TeV
even when the uplifting potential originates from a brane-localized source
located at the IR end of warped throat.
This scheme utilizes the holomorphic gauge kinetic function and non-perturbative
superpotential depending on both the dilaton superfield $S$ and the volume
modulus superfield $T$ whose axion components are periodic fields.
Following KKLT \cite{Kachru:2003aw}, we assume that $S$ is stabilized by flux
with a mass hierarchically heavier than the gravitino mass $m_{3/2}$, while $T$
is stabilized by non-perturbative superpotential with
$m_T\sim m_{3/2}\ln(M_{Pl}/m_{3/2})$.
In fact, such scheme has been studied recently in \cite{Abe:2005rx}, however the
possibility of $M_{\rm mir}\sim 1$ TeV has not been explored.

In mirage mediation, the Higgs mass parameter $B$ can be another source of fine
tuning since the conventional SUGRA mechanism to generate $\mu$ typically gives
$B\sim m_{3/2}\sim 8\pi^2 m_{\rm SUSY}$.
As we will see, the dilaton-modulus mixing in gauge kinetic function and
non-perturbative superpotential provides a non-perturbative mechanism to
generate $B\sim m_{\rm SUSY}$ in mirage or anomaly mediation scenario with
$m_{3/2}\sim 8\pi^2 m_{\rm SUSY}$.
Also, this mechanism for $B\sim m_{\rm SUSY}$ automatically gives a real $B/M_a$,
thus avoids the SUSY CP problem.

The mass pattern (I) and (II) differ by the values of $m_{H_d}$ and $B$, leading
to a significant difference in the Higgs spectrum and associated phenomenology.
A potential difficulty of the pattern (I) is that it requires a rather small
$B\sim M_Z/\tan\beta$, which might be difficult to be obtained even under a
mechanism to guarantee $B\sim m_{\rm SUSY}$. On the other hand,
the pattern (II) does not suffer from such difficulty, and predicts
$\tan\beta\sim \sqrt{8\pi^2}$ under a mechanism to give
$B\sim m_{\rm SUSY}\sim \sqrt{8\pi^2}M_Z$.
Although a rather extensive study of the mass pattern (I) has been performed in
\cite{Kitano:2005ew}, no detailed study of the mass pattern (II) has been made yet.
In the last part of this paper, we analyze the electroweak symmetry breaking and
various constraints from FCNC processes in both mass patterns of TeV scale mirage
mediation.

This paper is organized as follows. In section 2, we discuss the mirage mediation
resulting from a moduli stabilization set-up with dilaton-modulus mixing,
and also a non-perturbative mechanism to generate $B\sim m_{\rm SUSY}$ in mirage
mediation scenario.
We will present an explicit example which leads to the TeV scale mirage mediation
solving the little SUSY hierarchy problem while giving a desired size of
$B\sim m_{\rm SUSY}$.
In section 3, we discuss the electroweak symmetry breaking and the constraints
from FCNC processes for the SUSY mass patterns (I) and (II).
Section 4 is the conclusion.

\section{mirage mediation from a generalized moduli stabilization with
dilaton-modulus mixing}

In mirage mediation \cite{choi1}, soft terms receive comparable contributions from
anomaly mediation \cite{anomaly} and modulus mediation \cite{modulus}.
For the canonically normalized soft terms
\bea
{\cal L}_{\rm soft} &=&
-\frac{1}{2}M_a\lambda^a\lambda^a-\frac{1}{2}m_i^2|\phi^i|^2
-\frac{1}{6}A_{ijk}y_{ijk}\phi^i\phi^j\phi^k+{\rm h.c.},
\eea
where $\lambda^a$ are gauginos, $\phi^i$ are sfermions, $y_{ijk}$ are the
canonically normalized Yukawa couplings, the soft parameters at energy scale just
below the GUT scale $M_{GUT}$ are given by \cite{choi1}
\bea
\label{soft1}
M_a &=&
M_0 +\frac{b_a}{16\pi^2}g_{GUT}^2m_{3/2},
\nonumber \\
A_{ijk} &=&
\tilde{A}_{ijk}-\frac{1}{16\pi^2}(\gamma_i+\gamma_j+\gamma_k)m_{3/2},
\nonumber \\
m_i^2 &=&
\tilde{m}_i^2-\,\frac{1}{32\pi^2}\frac{d\gamma_i}{d\ln Q}m_{3/2}^2
+ \frac{1}{4\pi^2}
\left[\sum_{jk}\frac{1}{4}|y_{ijk}|^2\tilde{A}_{ijk}
-\sum_a g^2_aC^a_2(\phi^i)M_0\right]m_{3/2},
\eea
where $M_0$, $\tilde{A}_{ijk}$ and $\tilde{m}_i$ are the pure modulus-mediated
gaugino mass, trilinear $A$-parameters and sfermion masses which are generically
of the order of $m_{3/2}/8\pi^2$, and $Q$ denotes the renormalization scale.
Here $b_a$ and $\gamma_i$ are the one-loop beta function coefficients and the
anomalous dimension given by
\bea
b_a &=& -3{\rm tr}\left(T_a^2({\rm Adj})\right)+\sum_i
{\rm tr}\left(T^2_a(\phi^i)\right),
\nonumber \\
\gamma_i &=&2
\sum_a g_a^2C_2^a(\phi^i)-\frac{1}{2}\sum_{jk}|y_{ijk}|^2,
\eea
where the quadratic Casimir $C_2^a(\phi^i)=(N^2-1)/2N$ for a fundamental
representation $\phi^i$ of the gauge group $SU(N)$, $C_2^a(\phi^i)=q_i^2$ for the
$U(1)$ charge $q_i$ of $\phi^i$, and $\omega_{ij}=\sum_{kl}y_{ikl}y^*_{jkl}$ is
assumed to be diagonal.
Thus in our convention, $b_a$ and $\gamma_{H_u}$ of the MSSM are given by
\bea
b_a &=& (\frac{33}{5}, 1,-3),
\nonumber \\
\gamma_{H_u} &=& \frac{3}{2}g_2^2+\frac{1}{2}g_Y^2-3y_t^2,
\eea
where $g_2$ and $g_Y=\sqrt{3/5}\,g_1$ denote the $SU(2)_W$ and $U(1)_Y$ gauge
couplings.
For our later discussion, it is convenient to define
\bea
\alpha\equiv\frac{m_{3/2}}{M_0\ln(M_{Pl}/m_{3/2})},\quad
a_{ijk}\equiv \frac{\tilde{A}_{ijk}}{M_0}, \quad
c_i\equiv \frac{\tilde{m}_i^2}{M_0^2},
\eea
where $\alpha$ represents the anomaly to modulus mediation ratio, while $a_{ijk}$
and $c_i$ parameterize the pattern of the pure modulus -mediated soft masses.
As was noticed in \cite{choi1}, soft terms resulting from KKLT-type moduli
stabilization \cite{Kachru:2003aw} receive comparable contributions from both
the anomaly mediation and the modulus mediation, thereby $\alpha$, $a_{ijk}$ and
$c_i$ generically have the values of order unity.

Taking into account the 1-loop RG evolution, the above soft masses at $M_{GUT}$
lead to quite distinctive pattern of low energy soft masses which can be described
in terms of the mirage messenger scale \cite{Choi:2005uz}:
\bea
\label{mirage-scale}
M_{\rm mir} =
\frac{M_{GUT}}{(M_{Pl}/m_{3/2})^{\alpha/2}}.
\eea
The low energy gaugino masses are given by
\bea
M_a(Q) = M_0 \left[\,1-\frac{1}{8\pi^2}b_ag_a^2(Q)
\ln\left(\frac{M_{\rm mir}}{Q}\right)\,\right]
= \frac{g_a^2(Q)}{g_a^2(M_{\rm mir})}M_0,
\eea
showing that the gaugino masses are unified at $M_{\rm mir}$, while the gauge
couplings are unified at $M_{GUT}$.
The low energy values of $A_{ijk}$ and $m_i^2$ generically depend on the
associated Yukawa couplings $y_{ijk}$.
However if $y_{ijk}$ are small enough or if
\bea
\label{con1}
a_{ijk}=c_i+c_j+c_k=1,
\eea
their low energy values are given by \cite{Choi:2005uz}
\bea
\label{mirage-solution}
A_{ijk}(Q) &=& M_0\left[\,a_{ijk}+
\frac{1}{8\pi^2}(\gamma_i(Q) + \gamma_j(Q)+\gamma_k(Q))
\ln\left(\frac{M_{\rm mir}}{Q}\right)\,\right],
\nonumber \\
m_i^2(Q) &=& M_0^2\left[\,c_i-\frac{1}{8\pi^2}Y_i\left(
\sum_jc_jY_j\right)g^2_Y(Q)\ln\left(\frac{M_{GUT}}{Q}\right)\right.
\nonumber \\
&& \hspace{1cm} +\,
\left.\frac{1}{4\pi^2}\left\{
\gamma_i(Q)-\frac{1}{2}\frac{d\gamma_i(Q)}{d\ln Q}\ln\left(
\frac{M_{\rm mir}}{Q}\right)\right\}
\ln\left( \frac{M_{\rm mir}}{Q}\right)\,\right],
\eea
where $Y_i$ is the $U(1)_Y$ charge of $\phi^i$. Quite often,
the modulus-mediated squark and slepton masses have a common value, i.e.
$c_{\tilde{q}}=c_{\tilde{\ell}}$.
Then, according to the above expression of low energy sfermion mass, the 1st and
2nd generation squark and slepton masses are unified again at $M_{\rm mir}$.

A TeV scale mirage mediation  can provide a natural solution to the little SUSY
hierarchy problem \cite{Choi:2005hd,Kitano:2005wc}.
If $\alpha=2$ and also the conditions of (\ref{con1}) are satisfied for the top
quark Yukawa coupling, $M_{\rm mir}$ is of the order of 1 TeV and the troublesome
RG running of $m_{H_u}^2$ is nearly cancelled by the anomaly mediation effect.
Explicitly, we find
\bea
m_{H_u}^2(M_Z) &=& M_0^2\left[\,c_{H_u}- 0.026\sum_i c_iY_i
-\frac{3}{4\pi^2}y_t^2
\ln\left(\frac{M_{\rm mir}}{m_{\tilde{t}}}\right)
+ {\cal O}\left( \frac{1}{4\pi^2}\right)\,\right]
\nonumber \\
&=& c_{H_u}M_0^2 +{\cal O}\left(\frac{M_0^2}{4\pi^2}\right),
\eea
where $M_{\rm mir}\sim 1$ TeV. Related to the little SUSY hierarchy problem,
an attractive feature of mirage mediation arising from KKLT-type moduli
stabilization is that $\alpha$, $a_{ijk}$ and $c_i$ take {\it rational values}
(up to small corrections of ${\cal O}(1/4\pi^2)$) under suitable assumption.
Then by choosing the discrete parameters of the model in such a way to give
\bea
\label{tevmirage}
\alpha = 2, \quad c_{H_u}=0, \quad
a_{H_ut_Lt_R}=c_{\tilde{t}_L}+c_{\tilde{t}_R}=1,
\eea
one can naturally obtain the little hierarchy:
\bea
m_{H_u}^2(M_Z)\,\sim\,
\frac{m_{\rm SUSY}^2}{8\pi^2}\,\sim\, M_Z^2,
\eea
for which the correct EWSB can be achieved without any severe fine tuning of
parameters.
Here $m_{\rm SUSY}\sim M_0$ denotes generic superparticle masses including the
stop and gaugino masses.
The discrete parameter values of (\ref{tevmirage}) predict
\bea
M_{\tilde{g}}\simeq M_{\tilde{W}}\simeq M_{\tilde{B}} \simeq
A_t\simeq \sqrt{m_{\tilde{t}_L}^2+m_{\tilde{t}_R}^2}=
{\cal O}(\sqrt{8\pi^2}M_Z)
\eea
at low energy scales around 1 TeV, where $M_a$
($a=\tilde{g},\tilde{W},\tilde{B}$) are the MSSM gaugino masses, and
$m_{\tilde{t}_{L,R}}$ are the LL and RR stop masses.

Mirage mediation is a natural outcome of KKLT-type moduli
stabilization \cite{Kachru:2003aw} which can be described by 4D
effective action of the form \cite{choi1}: \bea \label{superspace}
\int d^4\theta \left[-3CC^*e^{-K/3} -C^2C^{*2}{\cal P}_{\rm
lift}\theta^2\bar{\theta}^2\right] + \left(\int d^2\theta \left[
\frac{1}{4}f_aW^{a\alpha}W^a_\alpha+C^3W\right] + {\rm h.c.}
\right), \eea where $C=C_0+F^C\theta^2$ is the chiral compensator
superfield, $K$ and $W$ are the K\"ahler potential and
superpotential, and ${\cal P}_{\rm lift}\theta^2\bar{\theta}^2$ is
the uplifting spurion operator induced by a SUSY breaking brane
which is assumed to be sequestered from the visible gauge and matter
superfields. After integrating out heavy moduli which are fixed by
fluxes, $K$ and $W$ appear to depend only on the light (volume)
modulus $T$ and the visible matter superfields $\Phi^i$: \bea K &=&
K_0(T+T^*)+Z_i(T+T^*)\Phi^{i*}\Phi^i,
\nonumber \\
W &=& W_0(T)+\frac{1}{6}\lambda_{ijk} \Phi^i\Phi^j\Phi^k.
\eea
Here we assume that the model possesses an axionic shift symmetry:
\bea
{\rm Im}(T)\rightarrow {\rm Im}(T)+\mbox{real constant},
\eea
which is broken by the non-perturbative term in $W_{0}$.
This ensures that the modulus K\"ahler potential $K_0$ and the matter K\"ahler
metric $Z_i$ depend only on the invariant combination $T+T^*$,
the holomorphic Yukawa couplings $\lambda_{ijk}$ are complex constants,
and finally $\partial_Tf_a$ are real constants.
These features eliminate the dangerous CP violating phases in soft terms deduced
from (\ref{superspace}) \cite{susycp}.

As long as the uplifting brane is sequestered from the visible gauge
and matter fields, its low energy consequence can be described by a
spurion operator \cite{choi1,pierce,hebecker} of the form \bea
\theta^2\bar{\theta}^2{\cal P}_{\rm lift}(T+T^*), \eea independently
of the detailed feature of SUSY breakdown.  The condition of nearly
vanishing cosmological constant requires \bea {\cal P}_{\rm
lift}={\cal O}(m_{3/2}^2M_{Pl}^2).\eea On the other hand, if  it is
induced by SUSY-breaking  at the IR end of warped throat as in the
scenario proposed by \cite{Kachru:2003aw}, which is the case of our
major concern,  ${\cal P}_{\rm lift}$ is red-shifted as \bea {\cal
P}_{\rm lift}\,\sim\, e^{4A}M_{Pl}^4,\eea where $e^{2A}\ll 1$ is the
metric warp factor at the end of throat, implying that $e^{2A}\sim
m_{3/2}/M_{Pl}$ in such scenario. Although it is possible that
uplifting is achieved by conventional $F$-term SUSY breaking which
is not necessarily sequestered from the volume modulus $T$
\cite{fuplifting}, here we focus on sequestered uplifting scenario
since the sequestering of visible sector is crucial for TeV scale
mirage mediation to solve the little SUSY hierarchy problem.

In the original KKLT compactification of type IIB string theory
\cite{Kachru:2003aw}, the uplifting operator is provided by an
anti-$D3$ brane stabilized at the IR end of warped throat, while the
CY volume modulus $T$ can be identified as a field living at the UV
end of throat \cite{hebecker1}.
 In such case,
$T$ is also sequestered from the uplifting brane, and thus ${\cal
P}_{\rm lift}$ is (approximately) independent of $T$ \cite{choi1}.
More detailed analysis of the modulus potential induced by anti-$D3$
\cite{giddings} and also the study of SUSY breaking transmitted
through warped throat \cite{kachru_sundrum} imply that
$\partial_T\ln{\cal P}_{\rm lift}={\cal O}(e^{4A})$ in the limit
that $T$ lives mostly in the unwarped region. As a result,
practically ${\cal P}_{\rm lift}$ can be regarded to be independent
of $T$ in scenarii that it originates from SUSY-breaking brane at
the IR end of warped throat.

The sequestering of visible matter, i.e. the suppression of the
dependence of ${\cal P}_{\rm lift}$ on the visible matter fields
$\Phi^i$: \bea \frac{\partial {\cal P}_{\rm
lift}}{\partial(\Phi^{i*}\Phi^j)}\,\ll\, m_{\rm SUSY}^2\,\sim\,
\left(\frac{m_{3/2}}{8\pi^2}\right)^2,\eea
 is crucial for mirage mediation to be able to give
$|m_{H_u}^2(M_Z)|\sim m_{\rm SUSY}^2/8\pi^2$ which would solve the
little SUSY hierarchy problem. It has been noticed in
\cite{dine,falkowski} that generically geometric separation alone
does not leads to such sequestering. In particular, for many
geometric background realized in string/$M$ theory, sizable contact
interaction (in $N=1$ superspace) between $\Phi^i$ and SUSY-breaking
field is induced by the exchange of bulk fields \cite{dine},
implying that a rather special type of geometric background is
required to realize sequestering.

On the other hand, studies of sequestering in some class of 4D CFT
\cite{conformalsequestering} and 5D warped geometry
\cite{luty_sundrum}, and also an operator analysis for SUSY-breaking
transmitted through warped throat \cite{choi3} suggest that
sequestering might be realized if the visible sector is separated
from SUSY-breaking brane by warped throat. Based on these
observations, sequestering of visible matter fields was  assumed in
the initial analysis of soft terms in KKLT set-up \cite{choi1}.
Recently, it has been argued in \cite{hebecker} that sizable contact
interaction  might be induced even for the case of warped throat by
the exchange of the throat isometry vector superfield.
More recently, this issue of sequestering in warped string
compactification has been examined in more detail
\cite{kachru_sundrum}, confirming that the desired sequestering can
be achieved easily when the visible brane and SUSY-breaking brane
are separated from each other by strongly warped throat. For
instance, it has been noticed that transmission of SUSY-breaking
through Klebanov-Strassler-type throat \cite{ks} leads to (in the
unit with $M_{Pl}=1$) \bea \frac{\partial {\cal P}_{\rm
lift}}{\partial(\Phi^{i*}\Phi^j)}\,\lesssim\, {\cal
O}(e^{\sqrt{28}A})\,=\,{\cal O}(e^{1.29A}m_{3/2}^2),\eea where we
have used  $e^{2A}\sim m_{3/2}/M_{Pl}$ for the metric warp factor.
The soft scalar masses of $\Phi^i$ resulting from this violation of
sequestering are given by \bea \delta m^2_{i\bar{j}}\,\lesssim\,
m_{3/2}^2\left(\frac{m_{3/2}}{M_{Pl}}\right)^{0.65} \,\sim\,
10^{-9}m_{3/2}^2,\eea which are small enough to be ignored compared
to the modulus and anomaly mediated scalar mass-squares of ${\cal
O}(m_{3/2}^2/(8\pi^2)^2)$.

The size of the violation of sequestering can differ for different
type of throat.  Generically, the warped sequestering scenario
discussed in \cite{kachru_sundrum} gives $\delta m^2_{i\bar{j}}\sim
e^{\gamma A}m_{3/2}^2$ with $\gamma={\cal O}(1)$ for the metric warp
factor which can be as small as $e^{2A}\sim m_{3/2}/M_{Pl}$, and
thereby the soft scalar masses of visible matter and Higgs fields
are dominated by the modulus and anomaly mediated contributions
given by (\ref{soft1}). In the following, we start with a set-up
including the case that ${\cal P}_{\rm lift}$ has a non-trivial
$T$-dependence as in Ref.
\cite{Choi:2005uz,Choi:2005hd,Kitano:2005wc}, while keeping that
${\cal P}_{\rm lift}$ is independent of the visible matter fields
$\Phi^i$. Later, we will focus on the specific case that ${\cal
P}_{\rm lift}$ is independent of both $T$ and $\Phi^i$.

In the Einstein frame, the modulus potential from (\ref{superspace}) takes the
form:
\bea
\label{moduluspotential}
V_{\rm TOT}
= e^{K_0}
\left[(\partial_T\partial_{\bar{T}}K_0)^{-1}|D_TW_{0}|^2-3|W_{0}|^2\right]
+ V_{\rm lift},
\eea
where $D_TW_0=\partial_TW_0+W_0\partial_TK_0$ and the uplifting potential is
given by
\bea
\label{uplift1}
V_{\rm lift}=
e^{2K_0/3}{\cal P}_{\rm lift}.
\eea
The superspace lagrangian density (\ref{superspace}) also determines the auxiliary
components of $C$ and $T$ as
\bea
\label{approx-F}
\frac{F^C}{C_0} &=& \frac{1}{3}\partial_TK_0F^T+m_{3/2}^*,
\nonumber \\
F^T &=& -e^{K_0/2}\left(\partial_T\partial_{\bar T}K_0\right)^{-1}
\left(D_TW_{0}\right)^*,
\eea
where $m_{3/2}= e^{K_0/2}W_0$.
For the minimal KKLT set-up with
\bea
\label{minimal}
f_a=T, \quad W_0=w_0-Ae^{-aT},
\eea
where $w_0$ is a hierarchically small constant of ${\cal O}(m_{3/2})$ and
$A={\cal O}(1)$ in the unit with $M_{Pl}=1$,
it is straightforward to compute the vacuum values of $T$ and $F^T$ by minimizing
the corresponding modulus potential (\ref{moduluspotential}) under the fine
tuning condition $\langle V_{\rm TOT}\rangle =0$\footnote{In fact, the correct
condition should be $\langle V_{\rm TOT}\rangle+\Delta V_{\rm TOT}=0$, where
$\Delta V_{\rm TOT}$ denotes the quantum correction to the classical vacuum
energy density $\langle V_{\rm TOT}\rangle$ \cite{ckn}.
This can alter the prediction of sfermion masses by an order of
$\Delta V_{\rm TOT}/M_{Pl}^2$. $\Delta V_{\rm TOT}$ is dominated by the
quadratically divergent one-loop corrections with the cutoff scale $\Lambda$,
i.e. $\Delta V_{\rm TOT}\sim N\Lambda^2 m_{\rm SUSY}^2/8\pi^2$, where $N$ is the
number of light superfields in 4D effective theory.
In KKLT-type moduli stabilization, the volume modulus is stabilized at a value
for which $\Lambda$ is comparable to $M_{GUT}$,
and then  $\Delta V_{\rm TOT}/M_{Pl}^2$ is small enough to be ignored.}.
At leading order in $\epsilon=1/\ln(M_{Pl}/m_{3/2})$, one finds
\cite{Choi:2005uz,choi_uni}:
\bea
\label{vev}
a T &=& \Big[ 1+{\cal O}(\epsilon)\Big]\ln (M_{Pl}/m_{3/2}),
\nonumber \\
M_0 &=& F^T\partial_T\ln({\rm Re}(f_a))=\frac{F^T}{T+T^*}
\nonumber \\
&=& \frac{m_{3/2}}{\ln(M_{Pl}/m_{3/2})}\left(1+\frac{3\partial_T
\ln({\cal P}_{\rm lift})}{2\partial_TK_0}
+ {\cal O}(\epsilon)\right),
\nonumber \\
\alpha &=& \frac{m_{3/2}}{M_0\ln(M_{Pl}/m_{3/2})}
= \left(1+\frac{3\partial_T\ln({\cal P}_{\rm lift})}{2\partial_T K_0}
+ {\cal O}(\epsilon)\right)^{-1}.
\eea
In order to get $\alpha=2$ giving $M_{\rm mir}\sim 1$ TeV within this minimal
set-up, one needs $\partial_T\ln({\cal P}_{\rm lift})/\partial_TK_0=-1/3$ as was
assumed in \cite{Choi:2005uz,Choi:2005hd,Kitano:2005wc}.
However, as was pointed out in \cite{pierce},
$\partial_T\ln({\cal P}_{\rm lift})/\partial_TK_0 < 0$  means that the uplifting
sector couples more strongly for a larger value of $T$, which makes it difficult
to give an extra-dimensional interpretation for ${\cal P}_{\rm lift}$.
Thus, in order to get $M_{\rm mir}\sim 1$ TeV in more plausible case with
$\partial_T\ln({\cal P}_{\rm lift})/\partial_TK_0 \geq 0$, one needs to modify
the minimal set-up given by (\ref{minimal}).

As was pointed out recently \cite{Abe:2005rx}, generalizing the gauge kinetic
functions as
\bea
\label{generalgauge}
f_a = k_aT+l_aS,
\eea
can give rise to a different value of the anomaly to modulus mediation ratio
$\alpha$ for a given form of ${\cal P}_{\rm lift}$, where $S$ is the dilaton
superfield and $T$ is the volume modulus superfield.
Such dilaton-modulus mixing in $f_a$ is {\it not} an unusual feature of string
compactification.
For instance, in heterotic string/M theory, for an appropriate normalization of
$S$ and $T$, one finds $l_a$ are positive rational numbers, while $k_a$ are
flux-induced rational number \cite{ck}: $k_a=\frac{1}{8\pi^2}
\int_{CY}J\wedge\left[{\rm tr}(F\wedge F)-\frac{1}{2}{\rm tr}(R\wedge R)\right]$,
where $J$, $F$ and $R$ are the K\"ahler, gauge and curvature 2-forms,
respectively.
Similar form of $f_a$ is obtained also in $D$-brane models of type II string
compactification.
For instance, the gauge kinetic function  on $D7$ branes wrapping a 4-cycle
$\Sigma_4$ is given by (\ref{generalgauge}) where  $k_a$ are integer-valued
wrapping number and $l_a$ are flux-induced rational number \cite{lust}:
$l_a=\frac{1}{8\pi^2} \int_{\Sigma_4}F\wedge F.$

The dilaton-modulus mixing in $f_a$ suggests that some non-perturbative terms in
the superpotential depend  on both $S$ and $T$ as
\bea
\label{generalsuper}
W_0=W_{\rm flux}(S,Z_\alpha)+W_{\rm np}=W_{\rm flux}(S,Z_\alpha)
+ \sum_I{A}_I(Z_\alpha)e^{-8\pi^2({k}_IT+{l}_IS)},
\eea
where $Z_\alpha$ are complex structure moduli stabilized by $W_{\rm flux}$
together with $S$, and ${A}_I$ generically have vacuum values of order unity.
Here $W_{\rm np}$ might be induced by hidden gaugino condensation or
string-theoretic instantons.
For a confining hidden $SU(N)$ gauge group with gauge kinetic function
$f_h=k_hT+l_hS$, gaugino condensation gives $W_{\rm np}\sim
e^{-8\pi^2 (k_hT+l_hS)/N}$.
Similarly, Euclidean action of some stringy instanton might be given by a linear
combination of $S$ and $T$, $S_{\rm ins}=8\pi^2(k_{\rm in}T+{l}_{\rm in}S)$,
thereby yielding $W_{\rm np}\sim e^{-8\pi^2({k}_{\rm in}T+{l}_{\rm in}S)}$.

An important feature of the gauge kinetic function (\ref{generalgauge}) and the
non-perturbative terms in (\ref{generalsuper}) which will be crucial for our
subsequent discussion is that
\bea
\label{rational}
\{ k_A/k_B,\, l_A/l_B\} =\mbox{rational numbers},
\eea
where $k_A=\{k_a,{k}_I\}$ and $l_A=\{l_a,{l}_I\}$.
Note that these ratios are determined by the topological or group theoretical
data of the underlying string compactification.
This feature can be easily understood  by noting that ${\rm Im}(S)$ and
${\rm Im}(T)$ are {\it periodic} axion fields, thus the coefficients $k_A$ and
$l_A$ should be quantized.
In the following, we discuss the mirage mediation resulting from the effective
SUGRA with the holomorphic gauge kinetic function (\ref{generalgauge}) and the
moduli superpotential (\ref{generalsuper}), and examine the possibility of
$M_{\rm mir}\sim 1$ TeV, i.e. $\alpha=2$, for a {\it sequestered}
uplifting function $\partial_T{\cal P}_{\rm lift}=0$.

Let us start with the usual KKLT assumption that $S$ and $Z_\alpha$ are fixed
by $W_{\rm flux}$ at $\langle S\rangle=S_0$ and
$\langle Z_\alpha\rangle=Z_{0\alpha}$ with a mass hierarchically heavier than
$m_{3/2}$ \cite{Kachru:2003aw}.
To be specific, we consider a model with the following form of the visible sector
gauge kinetic function and the moduli superpotential:
\bea
f_v &=& T+lS,
\nonumber \\
W_0 &=& W_{\rm flux}(S,Z_\alpha)+W_{\rm np}(S,Z_\alpha,T)\,=\,W_{\rm
flux}
-A_1e^{-8\pi^2(k_1T+{l}_1S)}, \eea where $A_{1}={\cal O}(1)$. Note
that we have chosen the normalization of $T$ for which
$k_A=(k_a,k_I)$ take rational values.
After integrating out the heavy $S$ and $Z_\alpha$,
 the effective gauge kinetic function and
modulus superpotential are given by \bea \label{eff} f^{({\rm
eff})}_v &=& T+lS_0,
\nonumber \\
W^{(\rm eff)}_{0} &=& \langle W_{\rm flux}\rangle +W_{\rm np}\,=\,
\langle W_{\rm flux}\rangle -A_1e^{-8\pi^2(k_1T+{l}_1S_0)}. \eea
Using $U(1)_R$ transformation and also the axionic shift of $T$, we
can always make $\langle W_{\rm flux}\rangle$ and
$A_1e^{-8\pi^2l_1S_0}$ real.

In the scheme under consideration, the requirement of nearly
vanishing cosmological constant leads to
 \bea {\cal P}_{\rm
lift}\,\sim\, \frac{|W_0^{(\rm eff)}|^2}{M_{Pl}^2}\,\sim\, m_{3/2}^2
M_{Pl}^2.\eea On the other hand, the volume modulus $T$ is
stabilized by $W^{(\rm eff)}_0= \langle W_{\rm flux}\rangle +W_{\rm
np}$ at a vacuum value yielding  $\langle W_{\rm np}\rangle\sim
\langle W_{\rm flux}\rangle/\ln(M_{Pl}/m_{3/2})$
\cite{Kachru:2003aw,choi1}. As a result, the flux-induced
superpotential is required to have a vacuum value \bea
\frac{|\langle W_{\rm flux}\rangle|^2}{M_{Pl}^2} \,\sim\,
\langle{\cal P}_{\rm lift}\rangle\eea  in order for the scheme to
admit the fine-tuning for nearly vanishing cosmological constant. In
case that ${\cal P}_{\rm lift}$ is induced by SUSY-breaking at the
IR end of warped throat as proposed in \cite{Kachru:2003aw}, one
finds \cite{GKP} \bea \frac{\langle{\cal P}_{\rm
lift}\rangle}{M_{Pl}^4}\,\sim\, e^{4A}\,\sim\, \exp\left[-\left(
\frac{-4\int_{\tilde{\Sigma}_3} H_3}{3\int_{\Sigma_3}
F_3}\right)8\pi^2 {\rm Re}(S_0)\right],\eea where $4\pi{\rm
Re}(S_0)=1/g_{st}$ for the string coupling $g_{st}$ whose self-dual
value is normalized to be unity, and $\int_{\Sigma_3}F_3$ and
$\int_{\tilde{\Sigma}_3}H_3$ denote the integer-valued RR and NS-NS
fluxes over the 3-cycle $\Sigma_3$ collapsing along the throat and
its dual 3-cycle $\tilde{\Sigma}_3$. To summarize, to achieve the
nearly vanishing cosmological constant, the flux-induced
superpotential is required to be tuned as (in the unit with
$M_{Pl}=1$) \bea |\langle W_{\rm flux}\rangle|\,\sim\,
\exp\left[-\left( \frac{-2\int_{\tilde{\Sigma}_3}
H_3}{3\int_{\Sigma_3} F_3}\right)8\pi^2 {\rm Re}(S_0)\right],\eea
thus can be parameterized as \bea \langle W_{\rm
flux}\rangle\,\equiv\, A_0e^{-8\pi^2l_0S_0}, \eea where
$l_0=-(2\int_{\tilde{\Sigma}_3}H_3)/(3\int_{\Sigma_3}F_3)$ is a {\it
positive rational} number of order unity and $A_0={\cal O}(1)$. As
we will see, this feature of the flux-induced superpotential makes
the vacuum value of ${\rm Re}(T)/l_0{\rm Re}(S)$ to be a {\it
rational} number (up to small corrections of ${\cal
O}(1/\ln(M_{Pl}/m_{3/2})$), which eventually yields the mirage
mediation parameters $\alpha$, $c_i$ and $a_{ijk}$ taking {\it
rational} values.  In the following, we will adopt this
parametrization of $\langle W_{\rm flux}\rangle$ while keeping in
mind that it does not originate from a non-perturbative dynamics,
but from the fine-tuning of the cosmological constant.

Minimizing the modulus potential (\ref{moduluspotential}) for \bea
W_0^{(\rm eff)}= \langle W_{\rm flux}\rangle
-A_1e^{-8\pi^2(k_1T+l_1S_0)}=A_0e^{-8\pi^2
l_0S_0}-A_1e^{-8\pi^2(k_1T+l_1S_0)}\eea and a generic uplifting
function ${\cal P}_{\rm lift}$, we find the vacuum values of $T$ and
$F^T$ are given by \bea \label{vev1} k_1 T &\simeq& (l_0-l_1)S_0+
\frac{1}{8\pi^2}\,\ln\left(\frac{-8\pi^2k_1}{\partial_TK_0}\frac{A_1}{A_0}\right),
\nonumber \\
\frac{F^T}{T+T^*} &\simeq&
\frac{l_0}{l_0-l_1}\frac{m_{3/2}}{\ln(M_{Pl}/m_{3/2})}
\left(1+\frac{3\partial_T\ln({\cal P}_{\rm
lift})}{2\partial_TK_0}\right). \eea On the other hand, the
phenomenologically favored $m_{3/2}\sim 10$ TeV and
$g_{GUT}^{-2}\simeq 2$ require \bea \label{stvev} 8\pi^2l_0{\rm
Re}(S_0) &\simeq& \ln(M_{Pl}/m_{3/2}) \sim 4\pi^2,
\nonumber \\
{\rm Re}(T)&+&l{\rm Re}(S_0)\,\simeq\, 2, \eea implying \bea 4
\,\lesssim\, \frac{l_0-l_1+k_1l}{k_1l_0} \,\lesssim\, 5 \eea when
the involved uncertainties are taken into account. The
modulus-mediated gaugino mass is given by \bea
\label{modulusgaugino} M_0 = F^T\partial_T\ln({\rm Re}(f_v)) =
\frac{F^T}{T+T^*} \left(\frac{l_0-l_1}{l_0-l_1+k_1l}\right), \eea
thus we find \bea \label{alpha} \alpha =
\frac{m_{3/2}}{M_0\ln(M_{Pl}/m_{3/2})} = \frac{l_0-l_1+k_1l}{l_0}
\left(1+\frac{3\partial_T\ln({\cal P}_{\rm lift})}{2\partial_TK_0}
\right)^{-1} \eea up to small corrections of the order of
$1/\ln(M_{Pl}/m_{3/2})\sim 1/4\pi^2$. Note that the $F$-component of
heavy dilaton $S$ is given by $F^S/S_0\sim m_{3/2}^2/m_S$, thus is
completely negligible since the dilaton mass $m_S$ is hierarchically
heavier than $m_{3/2}$.

The modulus-mediated $A$-parameters and sfermion masses are determined by the
following term in the superspace action (\ref{superspace}):
\bea
\int d^4\theta CC^* e^{-K_0/3}Z_i \Phi^{i*}\Phi^i,
\eea
where $K_0$ is the modulus K\"ahler potential and $Z_i$ is the matter K\"ahler
metric.
One then finds \cite{modulus}
\bea
\tilde{A}_{ijk}= a_{ijk}M_0 &=& F^T\partial_T\ln(e^{-K_0}Z_iZ_jZ_k),
\nonumber \\
\tilde{m}^2_i= c_iM_0^2 &=&
-|F^T|^2\partial_T\partial_{\bar{T}}\ln(e^{-K_0/3}Z_i). \eea In the
absence of dilaton-modulus mixing, $e^{-K_0/3}Z_i$ typically takes
the form: \bea e^{-K_0/3}Z_i=(T+T^*)^{n_i}, \eea where $n_i$ is a
rational number. The gauge flux leading to the modification of $f_a$
can modify the matter K\"ahler metric $Z_i$ also. For simplicity,
here we consider the case that the matter K\"ahler metric of the
visible sector is not affected by the involved dilaton-modulus
mixing, thereby $e^{-K_0/3}Z_i$ takes the above simple form.
 Then the
resulting $a_{ijk}$ and $c_i$ are  found to be \bea a_{ijk} &=&
(n_i+n_j+n_k)\left(\frac{l_0-l_1+k_1l}{l_0-l_1}\right),
\nonumber \\
c_{i} &=& n_i\left(\frac{l_0-l_1+k_1l}{l_0-l_1}\right)^2. \eea

In mirage mediation, the Higgs mass parameter $B$ can be another source of
fine tuning since the conventional SUGRA mechanism to generate $\mu$ generically
gives $B\sim 8\pi^2 m_{\rm SUSY}$ which is too large to give successful
electroweak symmetry breaking.
For instance, the Higgs bilinear terms in the K\"ahler and superpotential:
\bea
\label{mu}
\Delta K = \tilde\kappa(T+T^*)
H_uH_d+{\rm h.c.},\quad \Delta W=\tilde{\mu}(T)H_uH_d
\eea
give the canonically normalized Higgsino mass:
\bea
\mu = \mu_K+\mu_W=\frac{1}{\sqrt{Z_{H_u}Z_{H_d}}}
\left(m_{3/2}+F^{\bar{T}}\partial_{\bar{T}}\right)\tilde\kappa
+ \frac{1}{\sqrt{Z_{H_u}Z_{H_d}}}e^{K_0/2}\tilde{\mu},
\eea
and the canonically normalized holomorphic Higgs mass:
\bea
B\mu = -\Big[\,m_{3/2}^* + F^T\partial_T\ln(\tilde{\mu})
+ {\cal O}(F^T)\,\Big]\mu_W + \Big[\,m_{3/2}^*+{\cal O}(F^T)\,\Big]\mu_K,
\eea
where $Z_{H_u}$ and $Z_{H_d}$ are the K\"ahler metrics of $H_u$ and $H_d$,
respectively. Since $m_{3/2}\sim 8\pi^2 m_{\rm SUSY}$ in mirage mediation,
this shows that indeed $B$ is generically of ${\cal O}(8\pi^2 m_{\rm SUSY})$.

The moduli stabilization set-up discussed above provides a
non-perturbative mechanism to generate $B\sim m_{\rm SUSY}$ without
fine tuning. To obtain the desired size of $\mu$ and $B$, let us
assume that $\tilde{\kappa}=\tilde\mu=0$ in perturbation theory due
to a symmetry $G$ under which $H_uH_d$ has a non-trivial
transformation, however an exponentially small $\tilde\mu\sim
e^{-8\pi^2({k}_2T+l_2S_0)}$ is generated by non-perturbative effect
which breaks $G$: \bea \label{nonperturbativemu} \Delta W = A_2
e^{-8\pi^2({k}_2T+l_2S_0)}H_uH_d. \eea Adding the above
non-perturbative $\mu$-term to the modulus superpotential
(\ref{eff}), the total non-perturbative superpotential  of the model
is given by \bea \label{tot} W_{\rm TOT} &=& A_0e^{-8\pi^2l_0
S_0}-A_1e^{-8\pi^2(k_1T+l_1S_0)} +A_2e^{-8\pi^2(k_2T+l_2S_0)}H_uH_d,
\eea yielding \bea \label{mub} \mu &=&
\frac{e^{K_0/2}A_2e^{-8\pi^2(k_2 T+l_2 S_0)}}{\sqrt{Z_{H_u}Z_{H_d}}}
\,\sim\, A_2m_{3/2}^{N_\mu},
\nonumber \\
B&=&8\pi^2k_2 F^T-m_{3/2}+{\cal O}(F^T)
\nonumber \\
&=& \left[\,\frac{2k_2}{k_1} \left(1+\frac{3\partial_T \ln({\cal
P}_{\rm lift})}{2\partial_TK_0}\right) - 1 \,\right] m_{3/2}+{\cal
O}(F^T), \eea where \bea
N_\mu=\frac{k_2}{k_1}\frac{l_0-l_1}{l_0}+\frac{l_2}{l_0},\eea and we
have used the vacuum expectation values (\ref{vev1}) and
(\ref{stvev}) for the last expressions of $\mu$ and $B$. This result
shows that $B\sim m_{\rm SUSY}$ with a proper size of $\mu$ can be
obtained  by choosing the involved rational coefficients as \bea
\label{naturalmub} && \frac{k_2}{k_1} \,=\,
\frac{1}{2}\left(1+\frac{3\partial_T \ln({\cal P}_{\rm
lift})}{2\partial_TK_0}\right)^{-1},
\nonumber \\
&& \frac{k_2}{k_1}\frac{l_0-l_1}{l_0}+\frac{l_2}{l_0} \,=\, 1, \eea
and $A_2$ has a value of ${\cal O}(10^{-2})$ or of ${\cal
O}(10^{-3})$ depending upon the necessary value of $\mu$. Note that
$\partial_T\ln({\cal P}_{\rm lift})/\partial_TK_0$ is typically a
rational number for the volume modulus $T$, and $A_2$ can be
naturally small since the symmetry $G$ is restored in the limit
$A_2=0$.

It would be nice if $k_2=k_1$ and $l_2=l_1$, so that the
non-perturbative $\mu$-term $e^{-8\pi^2(k_2T+l_2S_0)}H_uH_d$ has the
same dynamical origin as the non-perturbative term
$e^{-8\pi^2(k_1T+l_1S_0)}$ which stabilizes $T$. However, in view of
the condition (\ref{naturalmub}), it is possible only when
$\partial_T\ln({\cal P}_{\rm lift})/\partial_TK_0=-1/3$
\cite{Choi:2005hd}, for which it is hard to give an
extra-dimensional interpretation to ${\cal P}_{\rm lift}$
\cite{pierce}. In more plausible case that $\partial_T\ln({\cal
P}_{\rm lift})/{\partial_TK_0}\geq 0$, these two terms can not have
the same origin. However still they can naturally have the same
order of magnitude by choosing the discrete parameters to satisfy
$\frac{k_2}{k_1}\frac{l_0-l_1}{l_0}+\frac{l_2}{l_0}=1$. Another
interesting feature of this mechanism to generate $\mu$ is that the
resulting $B$ is automatically real in the field basis that
$m_{3/2}$ and $F^T$ are real, thus avoids the SUSY CP problem, as a
consequence of the axionic shift symmetry of $T$ \cite{susycp}. In
the most interesting case that the uplifting brane is located at the
IR end of warped throat, thus is sequestered from the volume modulus
$T$, i.e. $\partial_T{\cal P}_{\rm lift}=0$, the values of $k_2$ and
$l_2$ which give $\mu\sim A_2m_{3/2}$ and $B\sim m_{\rm SUSY}$ are
\bea k_2 \,=\, \frac{1}{2}\,k_1, \quad l_2 \,=\,
\frac{1}{2}\,(l_0+l_1). \eea

The non-perturbative $\mu$-term (\ref{nonperturbativemu}) can be generated by a
confining hidden $SU(N_c)$ gauge interaction with $N_f$ flavors of hidden quarks
$Q_h+Q_h^c$ and a singlet $X$.
As a specific example, let us consider a hidden sector with $G=Z_3$ symmetry under
which
\bea
X \rightarrow e^{i2\pi/3}X, \quad
H_uH_d\rightarrow e^{i2\pi/3}H_uH_d, \quad Q_hQ_h^c \rightarrow
e^{-i2\pi/3}Q_hQ_h^c.
\eea
Up to ignoring irrelevant higher dimensional operators, the hidden gauge kinetic
function and superpotential invariant under $Z_3$ are given by
\bea
f_h &=& k_h T+l_h S_0,
\nonumber \\
W_h &=& \lambda_1 X^3+\lambda_2 XQ_hQ_h^c +
h_1Q_hQ_h^cH_uH_d+h_2X^2H_uH_d. \eea Note that $Z_3$ forbids a bare
$H_uH_d$ term in gauge kinetic functions, K\"ahler potential and
superpotential. The $Z_3$ symmetry is anomalous under the hidden
$SU(N_c)$ gauge interaction, thus is broken by the non-perturbative
Affleck-Dine-Seiberg superpotential \cite{ads} \bea W_{\rm ADS}=
(N_c-N_f) \left(\frac{e^{-8\pi^2 f_h}}{{\rm
det}(Q_hQ_h^c)}\right)^{1/(N_c-N_f)}. \eea Then, after integrating
out the confining hidden sector while including the effect of
$W_{\rm ADS}$, one finds the following effective superpotential:
\bea W_h^{(\rm eff)} = A_3e^{-12\pi^2(k_2 T+l_2 S_0)} +
A_2e^{-8\pi^2({k}_2 T+l_2S_0)}H_uH_d, \eea where \bea k_2 =
\frac{2k_h}{3N_c-N_f}, \quad l_2=\frac{2l_h}{3N_c-N_f}. \eea Since
$e^{-12\pi^2(k_2 T+l_2 S_0)}\sim m_{3/2}^{3/2}$ in the unit
$M_{Pl}=1$ for the rational coefficients (\ref{naturalmub}), the
first term of $W_h^{(\rm eff)}$ can be safely ignored.

Adding the above $W_h^{(\rm eff)}$ to (\ref{eff}), we obtain the
total superpotential: \bea W_{\rm TOT} &=& A_0e^{-8\pi^2
l_0S_0}-A_1e^{-8\pi^2(k_1T+l_1S_0)} +
A_2e^{-8\pi^2(k_2T+l_2S_0)}H_uH_d. \eea Let us recall that the first
term in $W_{\rm TOT}$ corresponds to the flux-induced superpotential
parameterized as $\langle W_{\rm flux}\rangle\,\equiv\, A_0
e^{-8\pi^2 l_0S_0}$, which reflects the fine-tuning required for
nearly vanishing cosmological constant for an exponentially
red-shifted uplifting operator ${\cal P}_{\rm lift}\sim e^{-16\pi^2
l_0{\rm Re}(S_0)}$. The second term $e^{-8\pi^2(k_1T+l_1S_0)}$ might
be induced by $D3$ brane instanton or $D7$ brane gaugino
condensation with $f_{D7}\propto k_1T+l_1S_0$. It should be stressed
that although each of the three terms in $W_{\rm TOT}$ has a
different origin, they naturally have the same order of magnitudes.
Independently of the value of $l_1$, $T$ is stabilized at a vacuum
value making the first and second terms comparable to each other. As
for the $\mu$-term, we could get $\mu\sim A_2 m_{3/2}$ and $B\sim
m_{3/2}/8\pi^2$ by assuming that the rational coefficients $k_2$ and
$l_2$ satisfy (\ref{naturalmub}).

So far, we have discussed generic mirage mediation resulting from moduli stabilization
with dilaton-modulus mixing.
Let us finally examine if this generalized set-up allows a TeV scale mirage mediation
solving the little hierarchy problem for the case that the uplifting function is
sequestered as $\partial_T{\cal P}_{\rm lift}=0$.
Here we just present a simple example giving the parameters in (\ref{tevmirage}).
The model is defined by
\bea
\label{model}
f_v &=& T,
\nonumber \\
W_{\rm TOT} &=& A_0e^{-8\pi^2 S_0} - A_1e^{-4\pi^2(T-2S_0)} +
A_2e^{-2\pi^2T}H_uH_d, \eea where the non-perturbative $\mu$-term is
induced by hidden $SU(N_c)$ gauge interaction with $f_h=T$, $N_c=3$
and $N_f=1$. The first term of $W_{\rm TOT}$ is assumed to be
induced by flux which admits the fine-tuning for nearly vanishing
cosmological constant for the uplifting function given by ${\cal
P}_{\rm lift}\sim e^{-16\pi^2 {\rm Re}(S_0)}$, where the exponential
suppression of ${\cal P}_{\rm lift}$ is due to the exponentially
small warp factor. The second term  of $W_{\rm TOT}$ might be
induced by string-theoretic instanton and/or additional hidden gauge
interaction with gauge kinetic function $\propto T-2S_0$. The
uplifting function is assumed to be sequestered from the volume
modulus $T$, which would be the case if it originates from a
SUSY-breaking brane at the IR end of warped throat, so that \bea
\partial_T{\cal P}_{\rm lift}=0.
\eea
The modulus K\"ahler potential and the K\"ahler metric of $H_u$, $t_{L}$ and $t_R$ are
chosen to be
\bea
\label{mirage}
K_0 &=& -3\ln(T+T^*),
\nonumber \\
e^{-K_0/3}Z_{H_u} &=& \mbox{constant},
\nonumber \\
e^{-K_0/3}Z_{t_L} &=& e^{-K_0/3}Z_{t_R}=(T+T^*)^{1/2}.
\eea
It is straightforward to see that this model gives the necessary TeV scale mirage
mediation parameters:
\bea
\alpha =2, \quad
c_{H_u}=0, \quad a_{H_ut_Lt_R}=c_{\tilde{t}_L}+c_{\tilde{t}_R}=1,
\eea
as well as
\bea
\mu \,\sim\, A_2 m_{3/2}, \quad B \,\sim\,
\frac{m_{3/2}}{8\pi^2}\,, \quad g_{GUT}^{-2} \,\simeq\, 2.
\eea
This model can give either the mass patterns (I) or (II) of (\ref{masspattern}),
depending upon the choice of $e^{-K_0/3}Z_{H_d}$ and the possibility of a further
suppression of $B$.

\section{sparticle spectrum and constraints from electroweak symmetry breaking
and FCNC}

\begin{figure}[t]
\begin{center}
\begin{minipage}{15cm}
\centerline{
{\hspace*{-.2cm}\psfig{figure=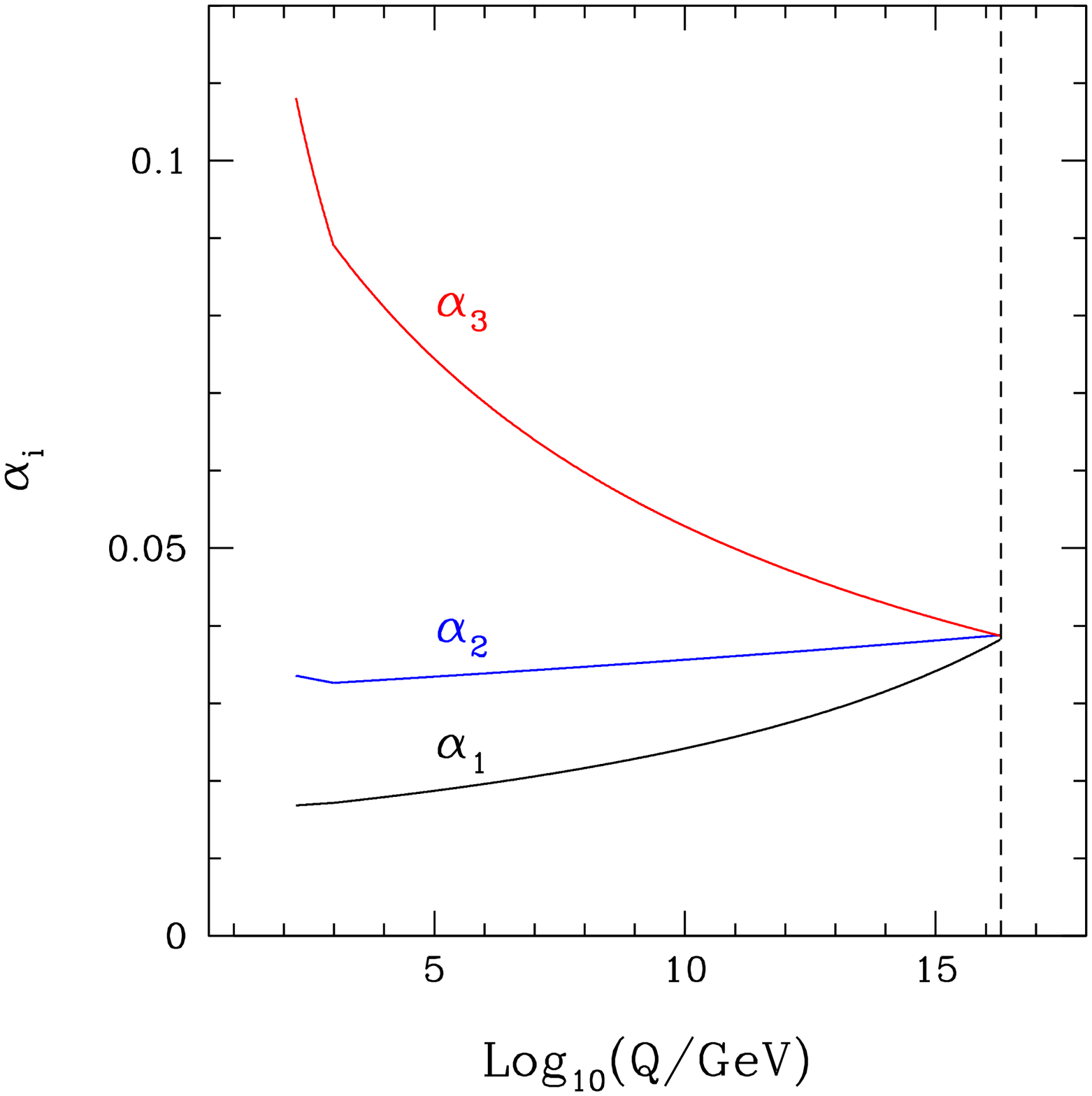,angle=0,width=7.5cm}}
{\hspace*{-.2cm}\psfig{figure=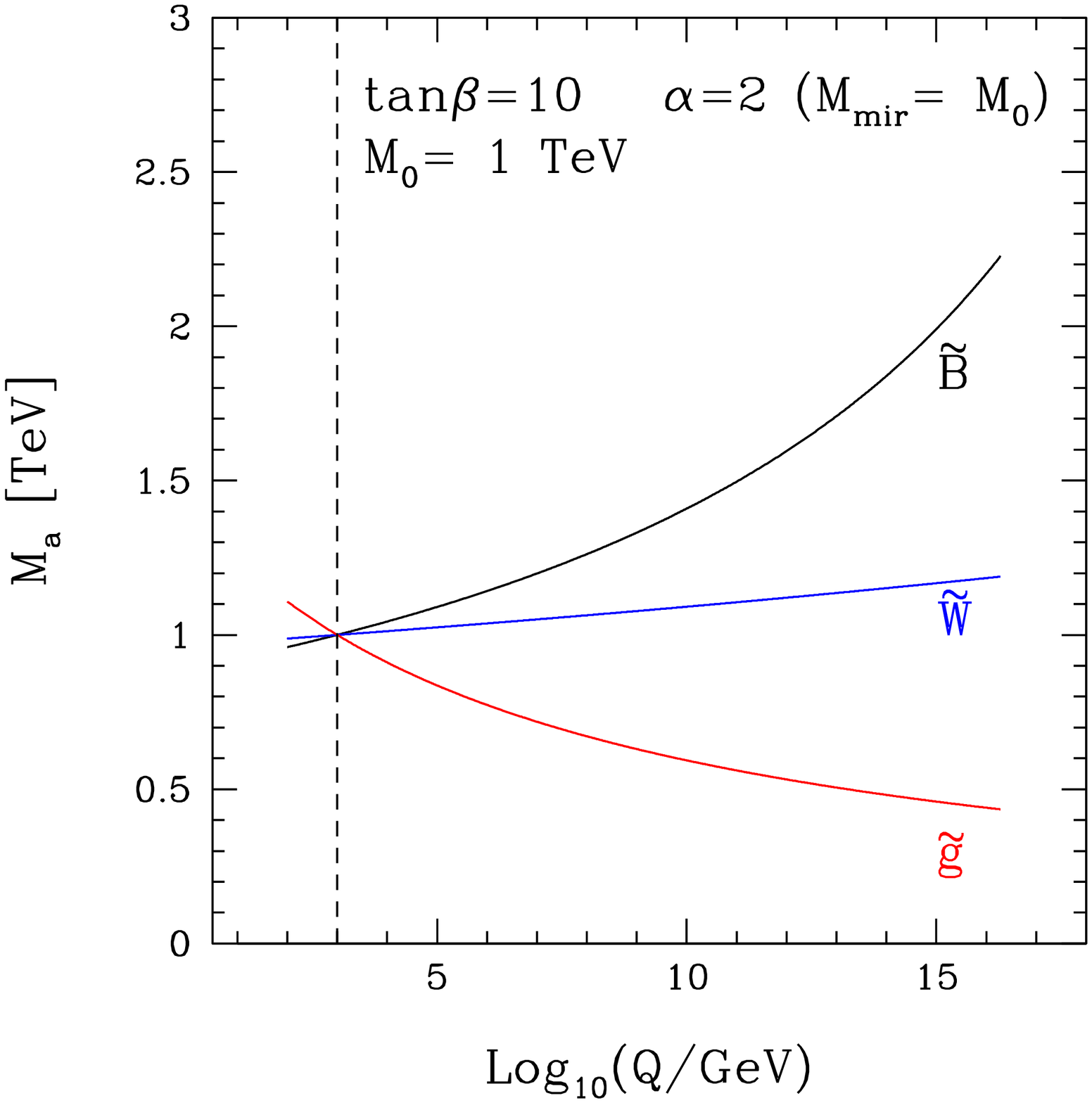,angle=0,width=7.5cm}}
}
\end{minipage}
\caption{Running of gauge couplings and gaugino masses in TeV
scale mirage mediation. \label{fig:gauge}}
\end{center}
\end{figure}

\begin{figure}[t]
\begin{center}
\begin{minipage}{15cm}
\centerline{
{\hspace*{-.2cm}\psfig{figure=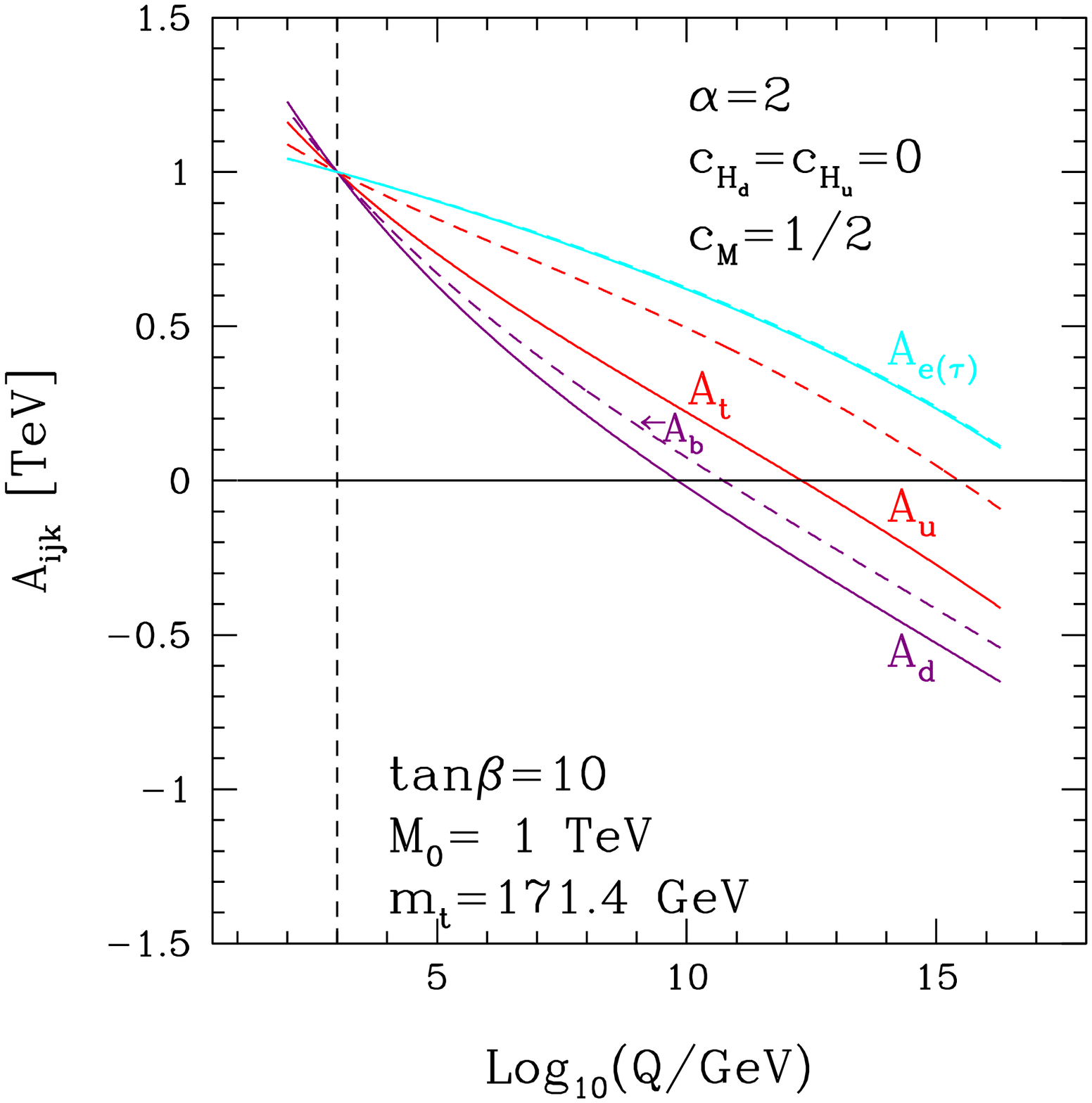,angle=0,width=7.5cm}}
{\hspace*{-.2cm}\psfig{figure=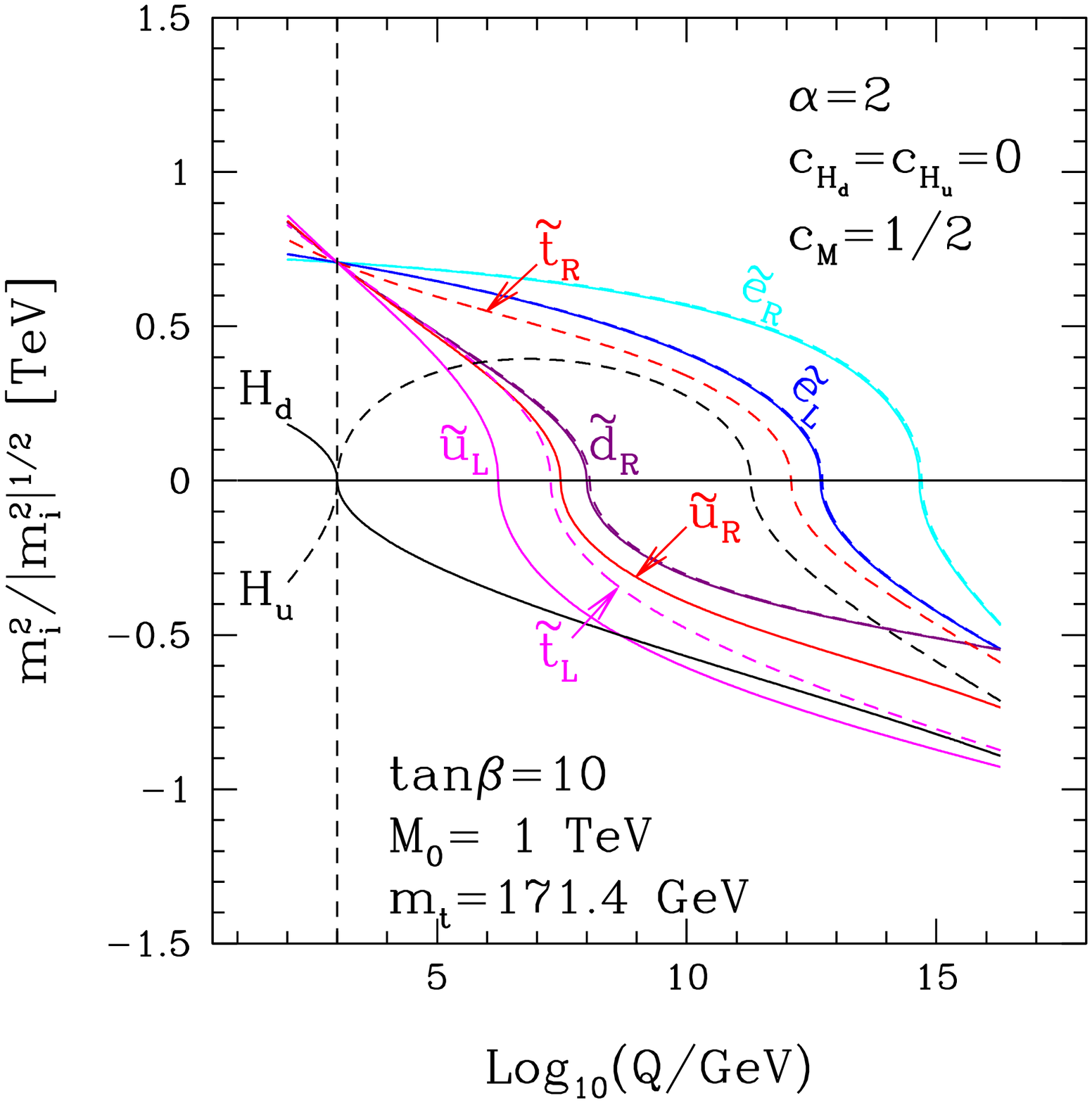,angle=0,width=7.5cm}}
}
\end{minipage}
\caption{Running of trilinear couplings and scalar masses leading to the mass pattern
(I).
\label{fig:pattern1}}
\end{center}
\end{figure}

\begin{figure}[t]
\begin{center}
\begin{minipage}{15cm}
\centerline{
{\hspace*{-.2cm}\psfig{figure=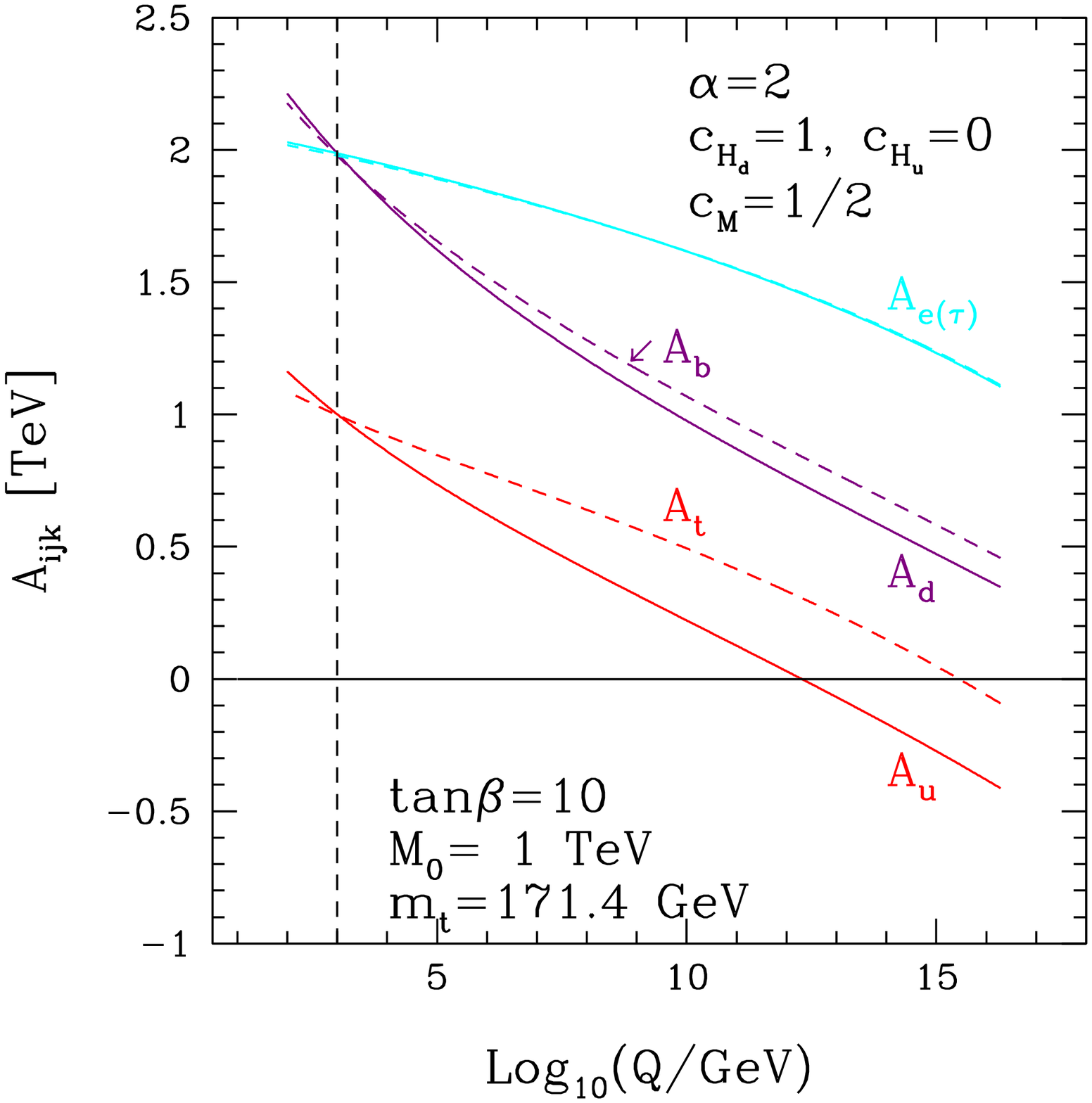,angle=0,width=7.5cm}}
{\hspace*{-.2cm}\psfig{figure=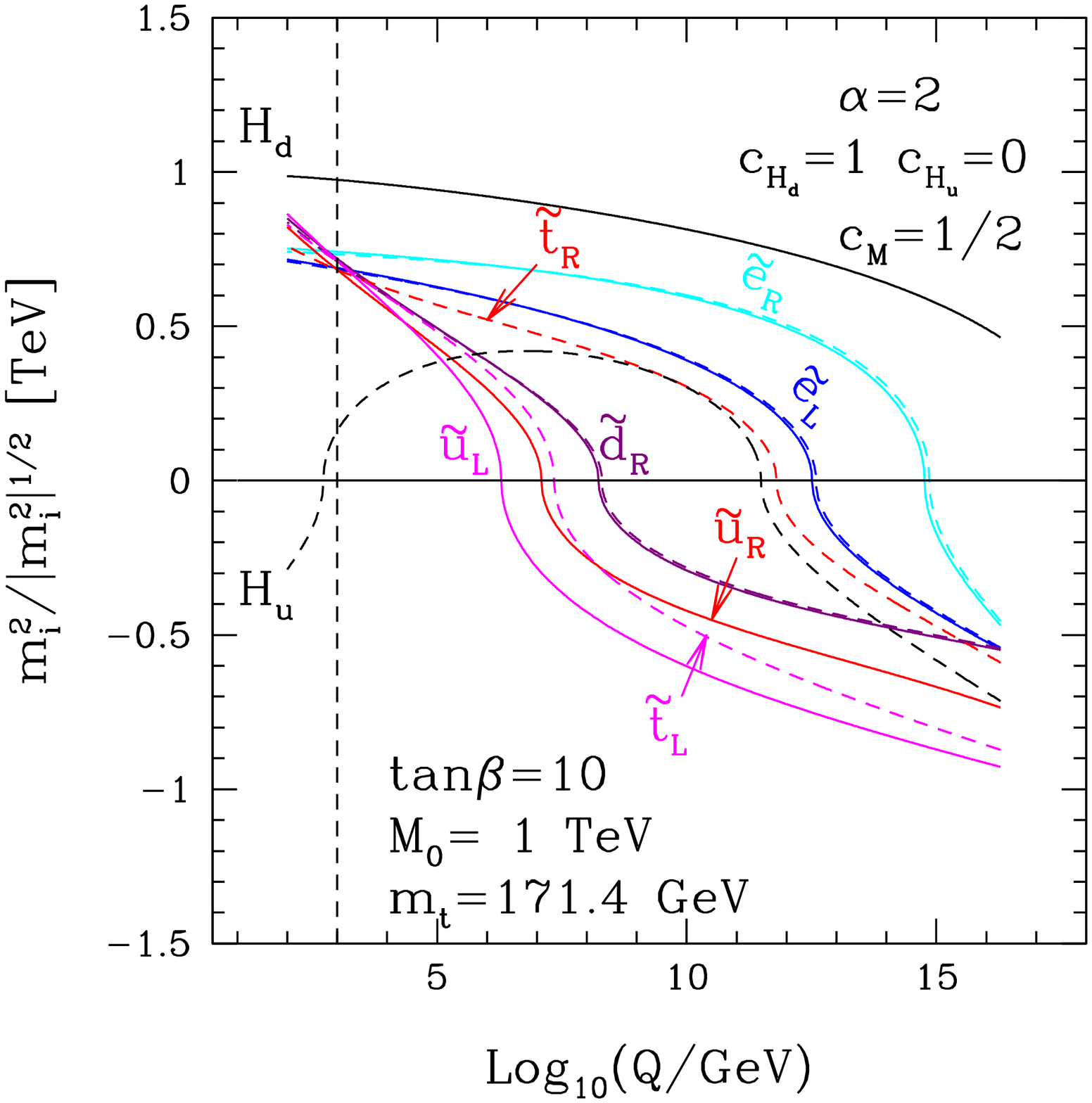,angle=0,width=7.5cm}}
}
\end{minipage}
\caption{Running of trilinear couplings and scalar masses leading to the mass pattern
(II).
\label{fig:pattern2}}
\end{center}
\end{figure}

In this section we discuss the low energy sparticle spectrum and the constraints
from electroweak symmetry breaking and FCNC processes in TeV scale mirage mediation
scenario.
The pattern of low energy sparticle masses  can be easily obtained by choosing
$M_{\rm mir}\sim 1$ TeV in the analytic solution (\ref{mirage-solution}).
In Fig.\ref{fig:gauge}, we show the running of gauge coupling constants and gaugino
masses.
Here we take $M_{\rm mir} = M_0 = 1$ TeV as a benchmark point.
Note that the gaugino masses are unified at $M_{\rm mir}$, while the gauge coupling
constants are  unified at $M_{GUT}\simeq 2.0\times 10^{16}$ GeV.
In Figs.\ref{fig:pattern1} and \ref{fig:pattern2}, we show the running of trilinear
couplings and scalar masses for the mass pattern (I) and (II), respectively.
We choose $c_i=1/2$ for all quark and lepton superfields, while $c_{H_u}=c_{H_d}=0$
for the mass pattern (I) and $c_{H_u}=0, c_{H_d}=1$ for the mass pattern (II).
As anticipated from (\ref{mirage-solution}), the trilinear couplings and scalar
masses are unified at $M_{\rm mir}$ while the Higgs soft masses cross zero for the
case of the mass pattern (I).
After taking into account the ambiguity in $M_{\rm mir}/M_0 ={\cal O}(1)$ and higher
order effects such as the threshold at $M_{GUT}$ and two-loop running, the model of
Fig.\ref{fig:pattern1} gives rise to the little hierarchy
$|m^2_{H_u, H_d}| \sim M_0^2/{8\pi^2}$ at $M_0\sim 1$ TeV.
For the model of Fig.\ref{fig:pattern2}, although the bottom Yukawa coupling and
the $U(1)_Y$ D-term contribution provide additional contribution to $m_{H_u}^2(M_0)$,
still a sufficient little hierarchy is realized for $m^2_{H_u}/M_0^2$, while
$m^2_{H_d}/M_0^2\sim 1$ in this case.

In TeV scale mirage mediation scenario, the squark/slepton
mass-squares renormalized at high energy scale, e.g. at a scale near
$M_{GUT}$, are negative as was noticed in \cite{adam}, while the
values at low energy scale below $10^6$ GeV are positive. Such
tachyonic {\it high energy} squark/slepton mass-squares might be
considered as a problematic feature of the model. However, as long
as the low energy squark/slepton mass-squares are positive, the
model has a correct color/charge preserving (but electroweak
symmetry breaking) vacuum which is a local minimum of the scalar
potential over the squark/slepton values $|\phi|\lesssim 10^6$ GeV.
On the other hand, tachyonic squark mass-squares at the RG point $Q
> 10^6$ GeV indicate that there might be a deeper color/charge
breaking (CCB) minimum or an unbounded from below (UFB) direction
\cite{munoz} at $|\phi|\gg 10^6$ GeV. In such situation, we need a
cosmological scenario in which our universe is settled down at the
correct vacuum with $\phi=0$.

In view of that the squarks and sleptons get large positive
mass-squares in the high temperature limit, it is a rather plausible
assumption \cite{kuzenko} that squark/slepton fields are settled
down at the color/charge preserving minimum after the inflation.
However, as was pointed out in  \cite{falk}, the early universe
might be trapped at the CCB minimum until when it becomes the global
minimum at low temperatures,
 depending upon the details of the model and also of the inflation
scenario. This should be avoided in order for TeV scale mirage
mediation to be viable. An examination of this issue is beyond of
this work as it requires an explicit scenario of early universe
inflation. We thus simply assume that TeV scale mirage mediation can
be combined with a successful early universe inflation leading to
squark/slepton vacuum values settled down at the color/charge
preserving local minimum.

Still we need to confirm that the color/charge preserving local
minimum is stable enough against the decay into CCB vacuum. It has
been noticed that the corresponding tunnelling rate is small enough,
i.e. less than the Hubble expansion rate, as long as the RG points
of vanishing squark/slepton mass-squares are all higher than $10^4$
GeV \cite{riotto,kuzenko}, which is satisfied safely by the TeV
scale mirage mediation scenario solving the little SUSY hierarchy
problem.

The analysis of electroweak symmetry breaking in TeV scale mirage mediation is more
involved because
\bea
\frac{dm_{H_u}^2}{d\ln Q}\,\sim\, \frac{m_{\rm SUSY}^2}{8\pi^2}\,\sim\, m_{H_u}^2,
\eea
around the TeV scale, and thus the running Higgs parameter
$m_{H_u}^2(Q)$ ($m_{H_d}^2(Q)$ also for the mass pattern (I)) is rather sensitive to
the RG point $Q$. To express the conditions for electroweak symmetry breaking in terms
of the RG-sensitive running parameters, one needs to include the Coleman-Weinberg
one-loop potential \cite{Coleman:1973jx, Gamberini:1989jw} which cancels the
$Q$-dependence coming from the running parameters.
This can be done efficiently \cite{Barger:1993gh} by replacing $m^2_{H_{d,u}}$ in the
electroweak symmetry breaking conditions derived from the RG-improved tree level Higgs
potential with
\bea
\label{replace}
\overline{m}^2_{H_{d,u}} = m^2_{H_{d,u}}-\frac{t_{d,u}}{\langle H_{d,u}^0 \rangle},
\eea
where the tadpoles $t_{d,u}$ are defined as
\bea
t_{d,u} = -\frac{1}{32\pi^2} Str\left[
\frac{\partial {\cal M}^2}{\partial \langle H_{d,u}^0 \rangle}
{\cal M}^2\left(\ln\left(\frac{{\cal M}^2}{Q^2}\right)-1\right)\right],
\eea
where $Str$ stands for the supertrace and ${\cal M}$ represents the full mass matrix
after $SU(2)_W\times U(1)_Y$ breaking.

Keeping this in mind, let us start with the RG-improved tree level scalar potential of
the neutral Higgs bosons in the MSSM:
\bea
\label{treepotential}
V &=& \left(m_{H_d}^2+|\mu|^2 \right) |H_d^0|^2
+ \left(m_{H_u}^2+|\mu|^2\right) |H_u^0|^2 - \left( B\mu H_d^0 H_u^0 +
{\rm h.c.} \right)
\nonumber\\
&&
+\, \frac{1}{8}\left(g^2_2+g_Y^2\right)\left({|H_d^0|}^2-{|H_u^0|}^2\right)^2.
\eea
This Higgs potential leads to $\langle H_{d,u}^0 \rangle \neq 0$ if the D-flat
direction is stable,
\bea
&& m_{H_d}^2+m_{H_u}^2+2|\mu|^2-2|B\mu| > 0,
\eea
and also the configuration $H_{d,u}^0=0$  is a saddle point,
\bea
&& \left(m_{H_d}^2+|\mu|^2 \right)\left(m_{H_u}^2+|\mu|^2 \right)
- |B\mu|^2 < 0.
\eea
At the minimum of the potential, $M_Z$ and
$\tan\beta=\langle H_u^0 \rangle / \langle H_d^0 \rangle$ are determined as
\bea
\label{eq:minimum}
&& \frac{M_Z^2}{2} = \frac{m^2_{H_d}-m^2_{H_u}\tan^2\beta}{\tan^2\beta-1}
- |\mu|^2,
\nonumber\\
&& \frac{1+\tan^2\beta}{\tan\beta}|B\mu| =
m^2_{H_d}+m^2_{H_u}+2|\mu|^2,
\eea
which correspond to the electroweak symmetry breaking conditions in the MSSM.

The second of the above electroweak symmetry breaking conditions has a solution only
when
\bea
&& \frac{m^2_{H_d}+m^2_{H_u}}{|B|^2}\le
\frac{1}{8}\left(\frac{1+\tan^2\beta}{\tan\beta}\right)^2.
\eea
We then find
\bea
|\mu| &=& \frac{1+\tan^2\beta}{4\tan\beta} |B|\left[ 1 \pm
\sqrt{1-8\left(\frac{\tan\beta}{1+\tan^2\beta}\right)^2
\frac{m^2_{H_d}+m^2_{H_u}}{|B|^2}  } \right],
\eea
where the minus sign is allowed only for $m^2_{H_d}+m^2_{H_u} \ge 0$.
In the expansion in powers of $1/\tan\beta$, these two solutions can be approximated
as
\bea
|\mu| &=&
\begin{cases}
\,\frac{\tan\beta}{2}\,|B|\left[1 +{\cal O}\left(\frac{1}{\tan^2\beta}\right)\right]&
\\
\,\frac{1}{\tan\beta}\frac{m^2_{H_d}+m^2_{H_u}}{|B|}\left[1 +{\cal O}
\left(\frac{1}{\tan^2\beta}\right)\right], & (m^2_{H_d}+m^2_{H_u}\ge 0).
\end{cases}
\eea
Combining with the first condition of (\ref{eq:minimum}), we can find the required
$|B|$ for given $m^2_{H_{d,u}}$ and $\tan\beta$.
The mass pattern (I) favors the first solution because $m^2_{H_d}+m^2_{H_u}$ tends to
be negative due to the large negative anomalous dimension of $H_u$.
On the other hand, the mass pattern (II) favors the second solution because the first
solution requires a too small $|B|$ to allow the solution itself.
This makes the two mass patterns behave in qualitatively different manner.
In particular, they require quite different size of $|B|$:
\bea
{\rm Pattern\, (I)}  &:& \quad |B| \,\simeq\,
\frac{2|\mu|}{\tan\beta} \,\sim\, \frac{M_Z}{\tan\beta},
\nonumber\\
{\rm Pattern\,(II)} &:& \quad |B| \,\simeq \,
\frac{1}{\tan\beta}\frac{m_{H_d}^2+m_{H_u}^2}{|\mu|} \,\sim\,
\frac{1}{\tan\beta}\frac{M_0^2}{M_Z}.
\eea

In mirage mediation, even when one has a mechanism to eliminate the contribution of
${\cal O}(m_{3/2})$ to $B$ as the one discussed in the previous section, it is hard
to control $|B|$ to make it significantly smaller than $M_0\sim m_{3/2}/4\pi^2$.
Note that generically $B$ can receive a contribution of ${\cal O}(m_{3/2}/8\pi^2)$ from
a threshold effect at the UV cutoff scale.
As a result, the mass pattern (I) might involve an additional fine-tuning to make $|B|$
as small as required.
On the other hand, the mass pattern (II) fits well to the natural prediction
$B\sim M_0$ which yields $\tan\beta\sim M_0/M_Z\sim \sqrt{8\pi^2}$.
In the following, we ignore this potential fine-tuning for the mass pattern (I),
and compare its phenomenological aspects with those of the mass pattern (II).

Our theoretical framework for mirage mediation can predict the soft parameters at TeV
with a precision of ${\cal O}(M_0/\sqrt{8\pi^2})$.
As a result, it provides only an order of magnitude prediction for the soft parameters
which have a size of ${\cal O}(M_0/\sqrt{8\pi^2})$ at TeV, i.e. $m_{H_u}$, $m_{H_d}$,
$B$ in the mass pattern (I) and $m_{H_u}$ in the mass pattern (II).
For these {\it small} parameters, we take a phenomenological approach treating them as
free input parameters defined at the electroweak scale within the range of
${\cal O}(M_0/\sqrt{8\pi^2})$ as suggested by the mirage mediation scheme.
To give a precise meaning to those input parameters, we define them at $Q=M_0/\sqrt{2}$
in the $\overline{DR}$ scheme \cite{Siegel:1979wq}.

The coupling constants and soft terms in the Higgs potential (\ref{treepotential}) are
running parameters and the result of analysis depends on the RG point $Q$ at which the
potential is minimized.
To deal with the Higgs parameters which have a size of ${\cal O}(M_0/\sqrt{8\pi^2})$,
we need to reduce this renormalization scale dependence by including the Coleman-Weinberg
one-loop effective potential \cite{Coleman:1973jx, Gamberini:1989jw}:
\bea
\Delta V_1 = \frac{1}{64\pi^2}Str\left[{\cal M}^4\left(
\ln\left(\frac{{\cal M}^2}{Q^2}\right)-\frac{3}{2}\right)\right].
\eea
This one-loop correction can be effectively included in (\ref{eq:minimum})
\cite{Barger:1993gh} by replacing $m^2_{H_{d,u}}$ with $\overline{m}^2_{H_{d,u}}$ defined
in (\ref{replace}).
Taking $c_{\tilde{q},\tilde{u},\tilde{d}}=1/2$ in (\ref{mirage-solution}), we obtain
\bea
\frac{t_{d}}{\langle H^0_{d} \rangle} &\approx& -\delta m^2_{H_{d}}
+ \delta B \mu\cdot {\rm sgn}(B\mu)\tan\beta,
\nonumber\\
\frac{t_{u}}{\langle H^0_{u} \rangle} &\approx& -\delta m^2_{H_{u}}
+ \delta B \mu\cdot {\rm sgn}(B\mu)\frac{1}{\tan\beta}\,,
\eea
where
\bea
\delta m^2_{H_{d,u}} &=&
-\frac{M_0^2}{8\pi^2}
\left[2 \gamma_{H_{d,u}}
\ln\left(\frac{M_0}{\sqrt{2}Q}\right)
+ \left(3 g^2_2+ g_Y^2\right) \ln(\sqrt{2})
+ \frac{1}{2}\left(3 y_{b,t}^2 -3 g^2_2 - g_Y^2\right) \right],
\nonumber \\
\delta B\mu &=&
\frac{\mu M_0}{8\pi^2}
\left[ \left(\gamma_{H_d}+\gamma_{H_u}\right)
\ln\left(\frac{M_0}{\sqrt{2}Q}\right)
+ \left(3g^2_2+g_Y^2\right)\ln(\sqrt{2})
- \frac{1}{2}\left(3 g^2_2 +g_Y^2\right) \right].
\eea
In the following numerical analysis, we use the electroweak symmetry breaking condition
(\ref{eq:minimum}) supplemented by the replacement
\bea
m_{H_{d,u}}^2\, \,\rightarrow\,\,
\bar{m}_{H_{d,u}}^2=m^2_{H_{d,u}}-\frac{t_{d,u}}{\langle H^0_{d,u}\rangle}\,,
\eea
which eliminates the sensitivity to the renormalization point $Q$ as the $Q$-dependence
from $m_{H_{d,u}}^2$ and $B$ is cancelled by the $Q$-dependence of
$t_{d,u}/\langle H^0_{d,u}\rangle$.
In this regard, the following estimate turns out to be useful:
\bea
\overline{m}^2_{H_u} \approx
\left.m^2_{H_u}\right|_{Q=\frac{M_0}{\sqrt{2}}} -0.95
\frac{M_0^2}{8\pi^2}.
\eea

In the model of the mass pattern (I), $m^2_{H_{d,u}}$ are free parameters of
${\cal O}(M_0^2/8\pi^2)$ at the weak scale.
If $\tan\beta$ is not too small, the first condition of (\ref{eq:minimum}) is
approximated as
\bea
\frac{M_Z^2}{2} \approx -\overline{m}^2_{H_u} -|\mu|^2
= -m^2_{H_u} + \frac{t_u}{\langle H^0_u \rangle} -|\mu|^2.
\nonumber
\eea
This leads to an upper bound of $m^2_{H_u}$,
\bea
m^2_{H_u} \lesssim -\frac{M_Z^2}{2} +\frac{t_u}{\langle H_u^0 \rangle},
\eea
which is saturated when $\mu=0$. Combining this with the second condition of
(\ref{eq:minimum}), we find
\bea
\label{eq:pseudo-Higgs mass}
m_A^2 &=& \frac{1+\tan^2\beta}{\tan\beta}|B\mu| \,\approx\,
\overline{m}^2_{H_d}-\overline{m}^2_{H_u}-M_Z^2
\nonumber\\
&\approx&
m^2_{H_d} - m^2_{H_u} - \frac{t_d}{\langle H_d^0 \rangle}
+ \frac{t_u}{\langle H_u^0 \rangle}-M_Z^2 \,\gtrsim\, 0,
\eea
where $m_A$ is the running pseudo-scalar Higgs mass which is of ${\cal O}(M_Z)$ in
this case.
In Fig.\ref{fig:finetuning1}, we show the parameter region leading to the correct
electroweak symmetry breaking on the planes of $(m_{H_u}^2, M_0)$ and
$(m_{H_u}^2, \tan\beta)$ for a benchmark scenario satisfying
$m_{H_d}^2/m_{H_u}^2 \approx \gamma_{H_d}/\gamma_{H_u}$.

In Fig.\ref{fig:finetuning2}, we present similar plots for the mass pattern (II).
In this case, $B$ is of ${\cal O}(M_0)$, which would be naturally achieved by the
non-perturbative mechanism discussed in the previous section,
and also $m^{2}_{H_d} \approx M_0^2$  and $m^2_{H_u}\sim M^2_0/8\pi^2$ at TeV under
the choice $c_{H_d}=1$ and $c_{H_u}=0$.
Then the electroweak symmetry breaking conditions lead to
\bea
\frac{M_Z^2}{2} &\approx&
\frac{\overline{m}^2_{H_d}}{\tan^2\beta}-\overline{m}^2_{H_u}-|\mu|^2
\approx
\left(1-\frac{\overline{m}^2_{H_d}}{|B|^2}\right)
\frac{\overline{m}^2_{H_d}}{\tan^2\beta}-\overline{m}^2_{H_u}.
\eea
Note that $m_A \simeq M_0$ in the case of the mass pattern (II).

Let us now estimate quantitatively the degree of fine-tuning for the electroweak
symmetry breaking.
Among the various possible measures of fine-tuning, we choose the sensitivity of $M_Z^2$
against a variation of the input parameter $\{a\}=\{ \mu^2, B, m^2_{H_{d,u}}\}$
\cite{Barbieri:1987fn}:
\bea
\Delta_a \equiv \frac{\partial \ln M_Z^2}{\partial \ln a}.
\eea
We then find
\bea
\Delta_{\mu^2} &=& -\frac{2|\mu|^2}{M^2_Z}
+ \frac{2\tan^2\beta}{(\tan^2\beta-1)^2}
\left(1 - \frac{4\tan\beta}{\tan^2\beta+1}\frac{|\mu|}{|B|} \right)
\left(1+\frac{\tan^2\beta+1}{\tan\beta}\frac{|B\mu|}{M^2_Z}\right),
\nonumber \\
\Delta_{|B|} &=&
\frac{4\tan^2\beta}{(\tan^2\beta-1)^2}
\left(1+\frac{\tan^2\beta+1}{\tan\beta}\frac{|B\mu|}{M^2_Z}\right),
\nonumber \\
\Delta_{m^2_{H_d}} &=& -\frac{2m^2_{H_d}}{M^2_Z}
\frac{1}{(\tan^2\beta-1)^2}
\left(\tan^2\beta+1+\frac{2\tan^3\beta}{\tan^2\beta+1}\frac{M^2_Z}{|B\mu|}\right),
\nonumber \\
\Delta_{m^2_{H_u}} &=& -\frac{2m^2_{H_u}}{M^2_Z}
\frac{\tan^2\beta}{(\tan^2\beta-1)^2}
\left(\tan^2\beta+1+\frac{2\tan\beta}{\tan^2\beta+1}\frac{M^2_Z}{|B\mu|}\right),
\eea
where we have taken account of the $\mu$ and $B$ dependence of $\tan\beta$.
For the mass pattern (I), $\Delta_a$ are simplified as
\bea
\Delta_{\mu^2} &=& -\frac{2|\mu|^2}{M_Z^2}
+ {\cal O}\left(\frac{1}{\tan^2\beta}\right),
\nonumber \\
\Delta_{|B|} &=&
\frac{4}{\tan^2\beta}\left(1+\frac{2|\mu|^2}{M_Z^2}\right)
+ {\cal O}\left(\frac{1}{\tan^4\beta}\right),
\nonumber \\
\Delta_{m^2_{H_d}} &=& -\frac{2m^2_{H_d}}{M_Z^2\tan^2\beta}
\left(1+\frac{M_Z^2}{|\mu|^2}\right)
+ {\cal O}\left(\frac{1}{\tan^4\beta}\right),
\nonumber \\
\Delta_{m^2_{H_u}} &=& -\frac{2m^2_{H_u}}{M_Z^2}
+{\cal O}\left(\frac{1}{\tan^2\beta}\right).
\eea
The above results show that $\Delta_{|B|}$ and $\Delta_{m^2_{H_d}}$ are subdominant
compared to $\Delta_{\mu^2}\sim \Delta_{m_{H_u}^2}$ if $|B|$ could be made to be
small enough to give $\tan^2\beta\sim M^2_Z/|B|^2\gg 1$.
Note that $\Delta_{|B|}$ measures the sensitivity of $M_Z^2$ to $|B|$ under the
assumption that $|B|$ is as small as $M_Z/\tan\beta$, not the degree of fine tuning
required to get such a small $|B|$.
$\Delta_{\mu^2}$ increases with $|\mu|$, but the degree of fine-tuning can be made
to be better than 10\%, i.e. $|\Delta_{\mu^2}^{-1}|>0.1$, for $|\mu| \lesssim 200$
GeV.
This is typically realized for a natural range of ${m}^2_{H_u}$ and $M_0$ as shown
in the left panel of  Fig.\ref{fig:finetuning1}.
We also plot in Fig.\ref{fig:finetuning1} the lightest Higgs mass using
{\it FeynHiggs1.2.2} \cite{FeynHiggs}.
The LEP bound on the physical Higgs boson mass, $m_{h^0}>114.4$ GeV, can be satisfied
with a fine-tuning of $\mu^2$ better  than 10\%.

For the mass pattern (II), the fine-tuning parameters are well approximated as
\bea
\Delta_{\mu^2} &=& \frac{2|\mu|^2}{M_Z^2}
\left(\frac{|B|^2}{m^2_{H_d}}-1\right)
+ {\cal O}\left(\frac{1}{\tan^2\beta}\right),
\nonumber\\
\Delta_{|B|} &=&
\frac{4m^2_{H_d}}{M_Z^2\tan^2\beta}
+ {\cal O}\left(\frac{1}{\tan^2\beta}\right),
\nonumber\\
\Delta_{m^2_{H_d}} &=&
-\frac{2m^2_{H_d}}{M_Z^2\tan^2\beta}
+ {\cal O}\left(\frac{1}{\tan^2\beta}\right),
\eea
where we have ignored the piece of  ${\cal O}(m_{H_u}^2/m_{H_d}^2)$, and the
expression for $\Delta_{m^2_{H_u}}$ is same as the one for the mass pattern (I).
In Fig.\ref{fig:finetuning2}, we show the parameter region for which
$|\Delta^{-1}_{\mu^2}|>0.1$ and $|\Delta^{-1}_{|B|}|>0.1$.
Note that $|\mu|$ can be significantly bigger than $M_Z$ while keeping
$|\Delta_{\mu^2}|={\cal O}(1)$ if $|B|^2\simeq m_{H_d}^2$.
This might open up an interesting possibility to raise $|\mu|$, thus raise the
Higgsino mass, without causing a serious fine tuning.

Let us finally discuss the constraints coming from various FCNC processes.
In the mass pattern (I), all Higgs bosons and higgsino have a light mass around a
few hundred GeV.
On the other hand, in the mass pattern (II), Higgs bosons other than the lightest
one have a mass close to 1 TeV,  while the higgsino  mass is around a few hundred
GeV.
In both cases, light particles can contribute to various FCNC processes through the
CKM-induced flavor mixing in the LR mixing masses.
This consideration results in some constraints on the model, particularly for large
$\tan\beta$ region, and provides an opportunity to test the model with future
experimental and theoretical improvements.
In Fig.\ref{fig:fcnc1} and Fig.\ref{fig:fcnc2}, we plot the constraints from
$b \to s\gamma$ for the mass pattern (I) and (II), respectively.
Current world average of the $b\rightarrow s\gamma$ branching ratio is given by
\cite{unknown:2006bi},
\bea
B(b \to s\gamma)_{E_\gamma>1.6 {\rm GeV}} = \left(3.55 \pm
0.24^{+0.09}_{-0.10}\pm 0.03\right) \times 10^{-4},
\eea
where $E_\gamma$ denotes the photon-energy cut.
Theoretical prediction of the SM is estimated as \cite{Gambino:2001ew}
\footnote{ Theoretical uncertainty quoted here is inherited mostly from input
parameters.
It has been argued that the photon-energy cut introduces another uncertainty
of similar size, which can be improved by perturbative calculation
\cite{Neubert:2004dd}.
Recent NNLO calculation  claims a central value 1.4 $\sigma$ lower than the experimental
world average \cite{Becher:2006pu}.
}
\bea
\label{bsgth}
B(b \to s\gamma)_{E_\gamma>1.6 {\rm GeV}}=3.57 \pm
0.30 \times 10^{-4}.
\eea
In Fig.\ref{fig:fcnc1} and Fig.\ref{fig:fcnc2}, we plot the 2-$\sigma$ range by
combining all the experimental and theoretical errors in quadrature:
\bea
\label{btosgamma}
2.75\times 10^{-4} < B(b \to
s\gamma)_{E_\gamma>1.6{\rm GeV}} < 4.35\times 10^{-4}.
\eea
In all plots, we choose a positive sign for $\mu$ for which the charged Higgs and
chargino contributions to $b\rightarrow s\gamma$ tend to cancel each other.
Negative $\mu$ gives a much stronger constraint due to the constructive interference.

In the mass pattern (I), both charged Higgs and chargino have a light mass,
and their contributions to $b\rightarrow s\gamma$ compete to each other depending on
the stop mass in the chargino contribution.
The left panel of Fig.\ref{fig:fcnc1} shows that a large fraction of $(m_{H_u}^2,M_0)$
leading to electroweak symmetry breaking gives a $b\rightarrow s\gamma$ branching
ratio within the allowed range (\ref{btosgamma}).
As the chargino contribution is enhanced by $\tan\beta$, the balance with the charged
Higgs contribution is somewhat sensitive to the value of $\tan\beta$.
In the right panel of Fig.\ref{fig:fcnc1} which is for the case of $M_0=1$ TeV,
the upper left (lower right) region with large (small) $\tan\beta$ is disfavored by
$b \to s\gamma$ due to the excessive chargino (charged Higgs) contribution
which gives a too small (large) branching ratio.

In the mass pattern (II), only the chargino contribution to $b\rightarrow s\gamma$ is
relevant.
Then the small $M_0~(\,\lesssim 800 \, {\rm GeV}\,) $ and large
$\tan\beta~(\,\gtrsim 15\,)$ regions in the left and right panels
of Fig.\ref{fig:fcnc2} are disfavored by giving a too small $b \to s\gamma$ branching
ratio.
In the right panel, the disfavored region quickly goes up and disappears if we
increase $M_0$.
Compared to the SM, the mass pattern (II) generically gives a smaller (larger)
branching ratio for $\mu>0$ ($\mu<0$).

In the mass pattern (I),  Higgs-mediated FCNC can give a sizable effect in large
$\tan\beta$ regime \cite{Huang:1998vb} since all Higgs bosons have a relatively light
mass.
We calculated $B_s \to \mu\mu$ rate \cite{Babu:1999hn} and also the double penguin
contribution to $\Delta m_{B_s}$ in $B_s$-$\overline{B}_s$ mixing \cite{Buras:2001mb}.
In the right panel of Fig.\ref{fig:fcnc1}, we plot contours for the branching ratio
of $B_s \to \mu\mu$.
The SM prediction is chirality suppressed,
$B_{SM}(B_s \to \mu\mu)=3.46 \times 10^{-9}$,
however this is not the case for supersymmetric contribution.
Current experimental bound, $B(B_s \to \mu\mu) < 1.0 \times 10^{-7}$, excludes a
region above $\tan\beta\simeq 30$ for $M_0=1$ TeV.
If the upper bound is improved to $1.0\times 10^{-8}$, the excluded region comes
down to $\tan\beta\simeq 20$.
The branching ratio reaches to $5\times 10^{-9}$ around $\tan\beta\simeq 10$.
However we note that it is rather unlikely that $\tan\beta\gtrsim 10$ in the mass
pattern (I) as it requires a very small $|B|\sim M_Z/\tan\beta$.

Recently a finite value of $\Delta m_{B_s}$ has been measured at Tevatron with an
unprecedented accuracy  \cite{Abazov:2006dm}:
\bea
\Delta m_{B_s} = 17.31^{+0.33}_{-0.18} \pm 0.07~ {\rm ps}^{-1}.
\eea
Double Higgs penguin contribution to $\Delta m_{B_s}$ can potentially cause a
significant deviation from the SM prediction in large $\tan\beta$ regime
\cite{Carena:2006ai,Foster:2006ze}.
We examined an impact of this measurement on the mass pattern (I), however could not
obtain a constraint stronger than the one from $B_s \to \mu\mu$.
This is mainly due to a large  ambiguity in  hadronic parameters which determine
the SM prediction.
This uncertainty can be reduced if we consider the ratio
$\Delta m_{B_s}/\Delta m_{B_d}$, however in this case the dependence on the poorly
known unitarity angle $\phi_3$ ($\gamma$) introduces another source of ambiguity
\cite{Ball:2006xx}.
Considering the accuracy of the measurement, future progress in the lattice
calculation of the involved hadronic parameters and also a precise determination of
the unitarity angle might make $\Delta m_{B_s}$ a strong probe for the mass
pattern (I).
For the mass pattern (II),  these Higgs mediated processes do not lead to any
significant deviation from the SM predictions.

Anomalous magnetic moment of the muon provides a powerful tool to test new physics
around the electroweak scale. Since the first report by BNL E821,
the SM prediction has been carefully examined and refined including the semi-empirical
estimation of hadronic vacuum polarization by dispersion relation and also
the model dependent estimation of hadronic light-by-light contribution.
See \cite{Davier:2003pw,Passera:2004bj} for recent progress.
Using data set from $e^+ e^-$ collisions for the hadronic vacuum polarization,
\cite{Passera:2004bj} quotes the SM prediction as
\bea
a_{\mu}^{\rm SM} = (11 659 184.5 \pm 6.9)\times 10^{-10},
\eea
while the latest experimental value is reported as \cite{Bennett:2004pv}
\bea
a_{\mu}^{\rm exp} = (11 659 208.0 \pm 6.0)\times 10^{-10}.
\eea
This amounts to $2.6\sigma$ deviation from the SM \footnote{The latest analysis claims
3.4 $\sigma$ deviation with a $4.1\times 10^{-10}$ larger central value
\cite{Hagiwara:2006jt}.}:
\bea
\Delta a_{\mu} \equiv a^{\rm exp}_\mu - a^{\rm SM}_{\mu} = (23.5\pm 9.1)
\times 10^{-10}
\eea
Analysis based on $\tau$ decays shows $0.7\sigma$ deviation, however this result is
still under debate due to the lack of full understanding of isospin-breaking effect.
Further theoretical and experimental effort will confirm or diminish the current
disagreement based on $e^+e^-$.

In the MSSM, charged Higgs contribution to the anomalous muon
magnetic moment is suppressed by small Yukawa couplings, and then
dominant contribution comes from chargino and neutralino loop
diagrams. In TeV scale mirage mediation scenario, the gaugino
contribution to $a_\mu$ is small as the bino and wino masses are
close to $M_0\sim 1$ TeV. The higgsino contribution is also small as
it is suppressed by small Yukawa couplings. As a result, $\Delta
a_\mu$ in TeV scale mirage mediation is significantly smaller than
the value obtained in the conventional scenarios which have light
gauginos and/or stau \cite{choi4}. In Fig.\ref{fig:fcnc1} and
\ref{fig:fcnc2}, we plot the SUSY contribution to the muon g-2 for
the mass pattern (I) and (II), respectively. Taking into account the
constraints from FCNC processes and the lightest higgs boson mass,
TeV scale mirage mediation scenario predicts \bea \Delta a_{\mu}
&\lesssim& 10\times 10^{-10} \quad \mbox{(the mass pattern (I))},
\nonumber \\
\Delta a_{\mu} &\lesssim& 5\times 10^{-10} \quad\,\,\,
\mbox{(the mass pattern (II))}.
\eea
If the discrepancy between the SM prediction based on $e^+e^-$ scattering and the
experimental measurement is confirmed with the current central value,
it can not be accommodated in the TeV scale mirage mediation set-up discussed here.
In this regard, an improvement of the theoretical and experimental errors on
$a_\mu$ will have a considerable impact on TeV scale mirage mediation scenario.

\section{Conclusion}

TeV scale mirage mediation has been proposed as a pattern of soft SUSY breaking
terms reducing the fine tuning for the electroweak symmetry breaking in the
MSSM \cite{Choi:2005hd,Kitano:2005wc},
thereby solving the little SUSY hierarchy problem.
The original proposal is based on a SUSY breaking uplifting potential which is
difficult to give an extra-dimensional interpretation \cite{pierce}.
In this paper, we note that the desired form of TeV scale mirage mediation can be
achieved within a moduli stabilization scheme which has a brane-localized
(sequestered) origin of the SUSY-breaking uplifting potential,
if the holomorphic gauge kinetic functions and non-perturbative superpotential
depend on both the dilaton superfield $S$ and the volume modulus superfield $T$.
We also propose a non-perturbative mechanism to generate the Higgs $B$ parameter
which has a desirable size $B\sim m_{\rm SUSY}\sim m_{3/2}/8\pi^2$ in mirage
mediation scheme.
An important feature of the scheme is that the axion components of $S$ and $T$
are periodic fields, therefore the coefficients of $S$ and $T$ in gauge kinetic
functions and non-perturbative superpotential can have discrete values only.
As in the case of KKLT moduli stabilization, $S$ is assumed to be stabilized by flux
with $m_S$ hierarchically heavier than $m_{3/2}$,
while $T$ is stabilized by non-perturbative effects yielding
$m_T\sim m_{3/2}\ln(M_{Pl}/m_{3/2})$.
Then, under a proper choice of the involved discrete parameters, the TeV scale
mirage mediation pattern of soft parameters solving the little SUSY hierarchy
problem can be obtained .

\begin{figure}[t]
\begin{center}
\begin{minipage}{15cm}
\centerline{
{\hspace*{-.2cm}\psfig{figure=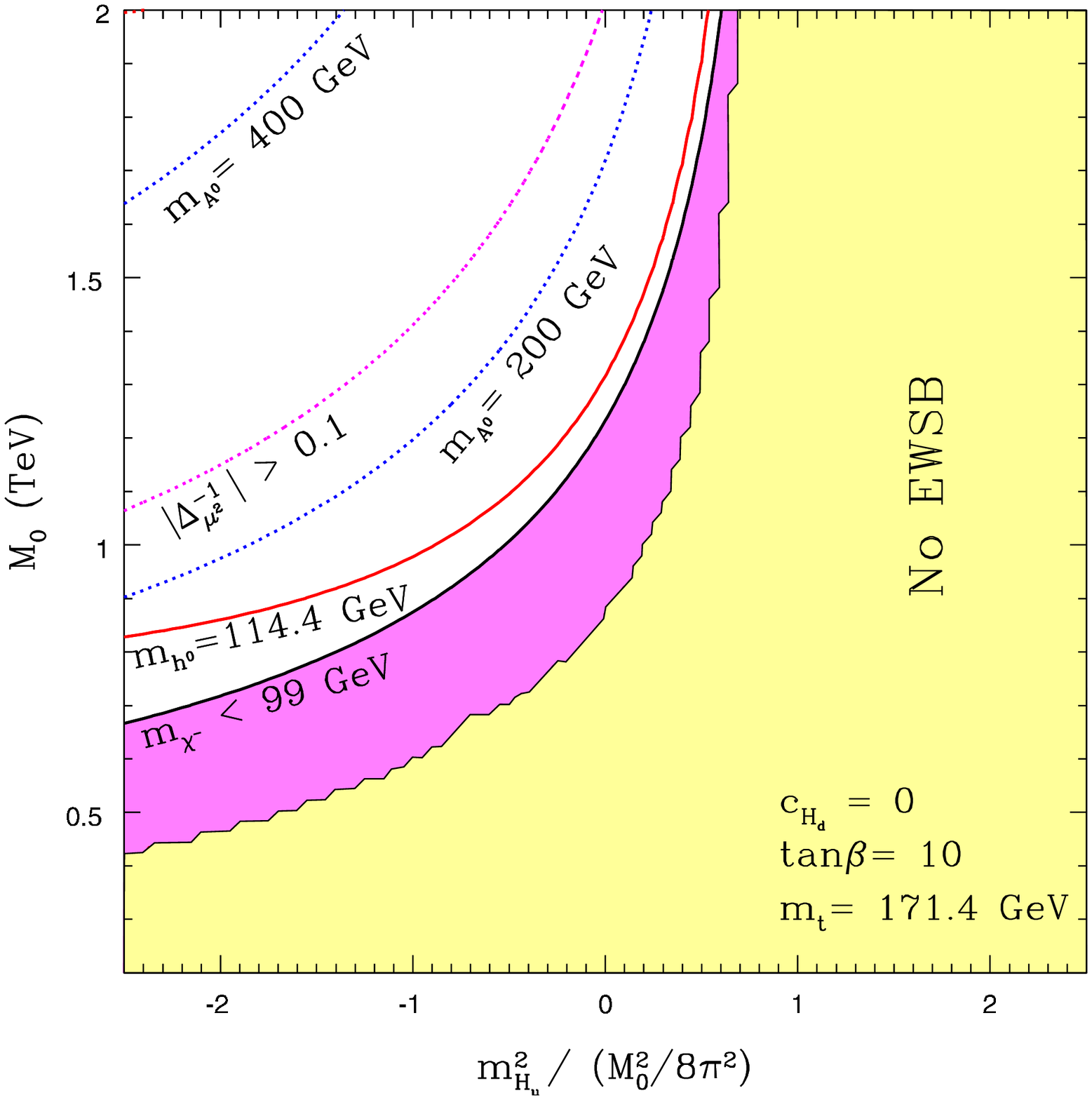,angle=0,width=7.5cm}}
{\hspace*{-.2cm}\psfig{figure=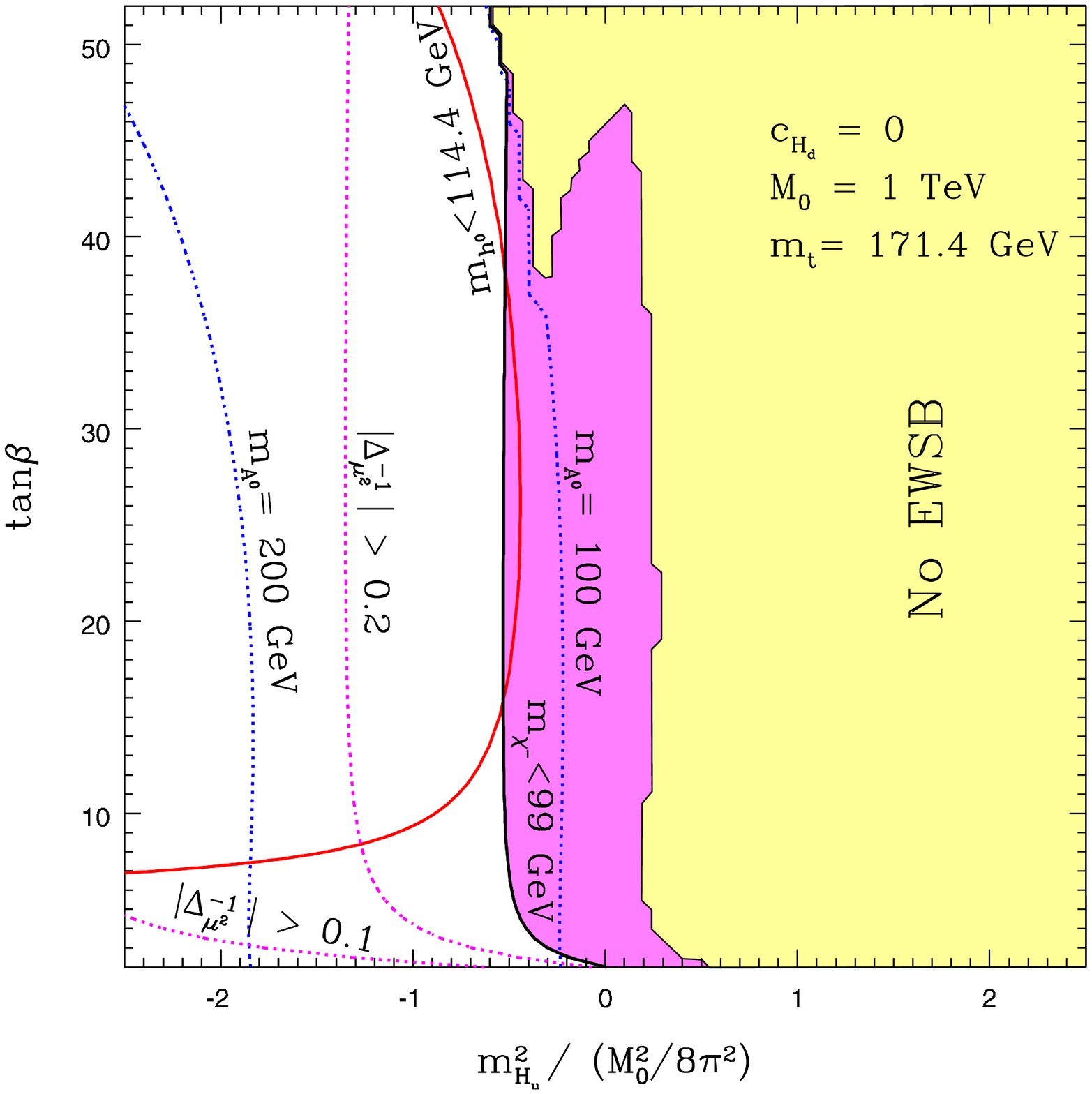,angle=0,width=7.5cm}}
}
\end{minipage}
\caption{Electroweak symmetry breaking, Higgs boson masses and the degree of
fine-tuning in the mass pattern (I).
\label{fig:finetuning1}}
\end{center}
\end{figure}

\begin{figure}[t]
\begin{center}
\begin{minipage}{15cm}
\centerline{
{\hspace*{-.2cm}\psfig{figure=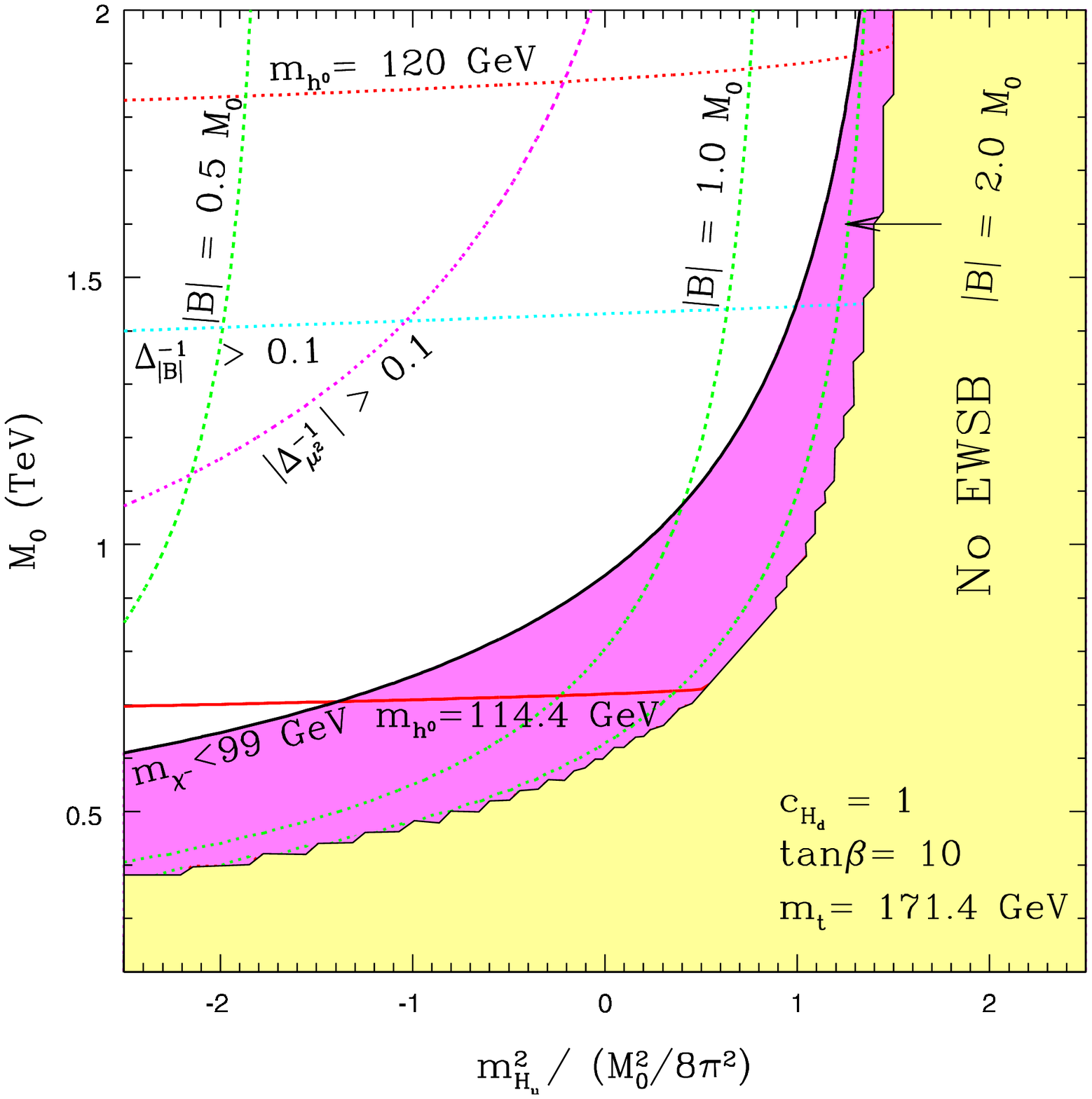,angle=0,width=7.5cm}}
{\hspace*{-.2cm}\psfig{figure=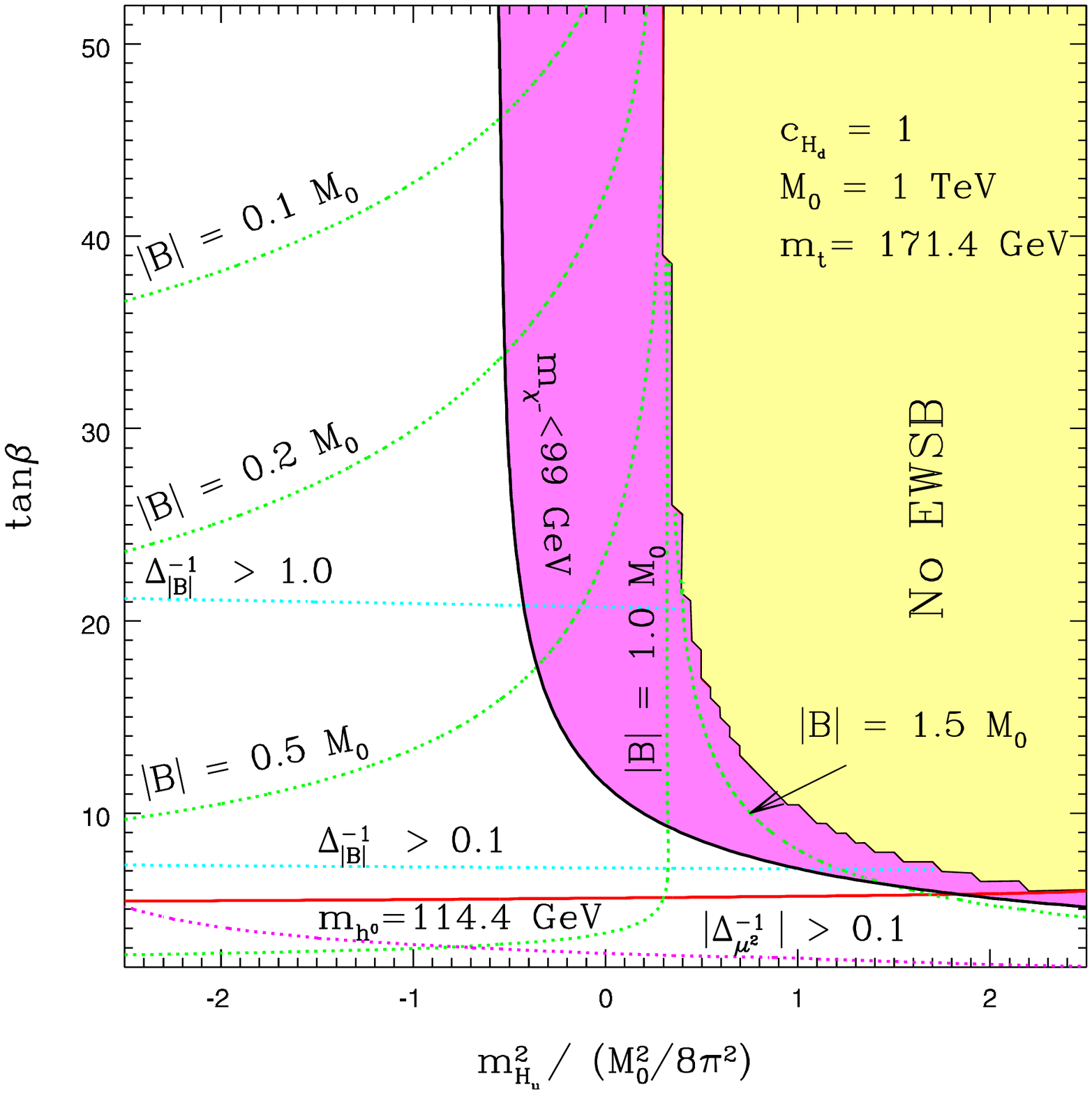,angle=0,width=7.5cm}}
}
\end{minipage}
\caption{Electroweak symmetry breaking, Higgs boson masses and the degree of
fine-tuning in the mass pattern (II).
\label{fig:finetuning2}}
\end{center}
\end{figure}

\begin{figure}[t]
\begin{center}
\begin{minipage}{15cm}
\centerline{
{\hspace*{-.2cm}\psfig{figure=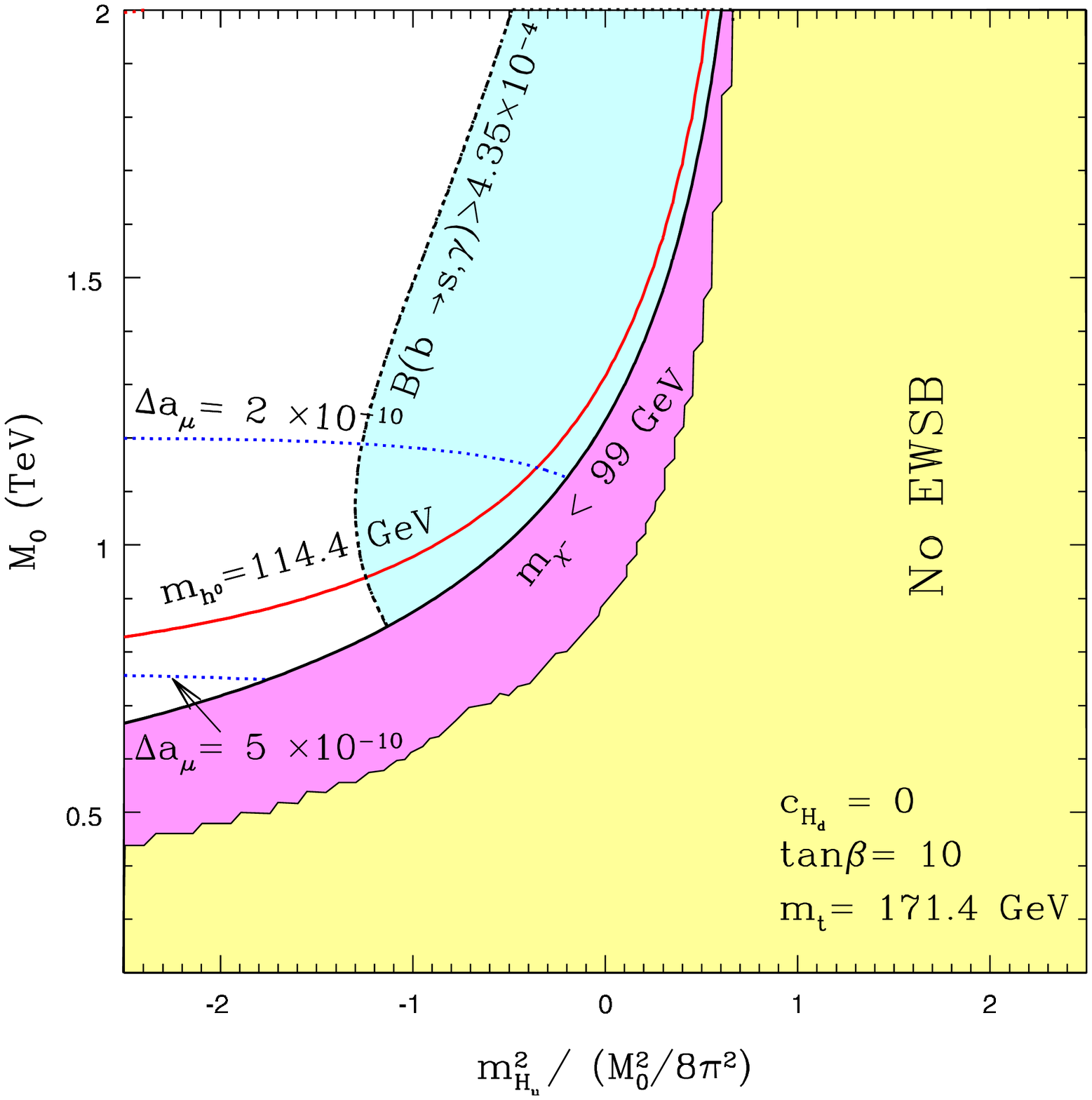,angle=0,width=7.5cm}}
{\hspace*{-.2cm}\psfig{figure=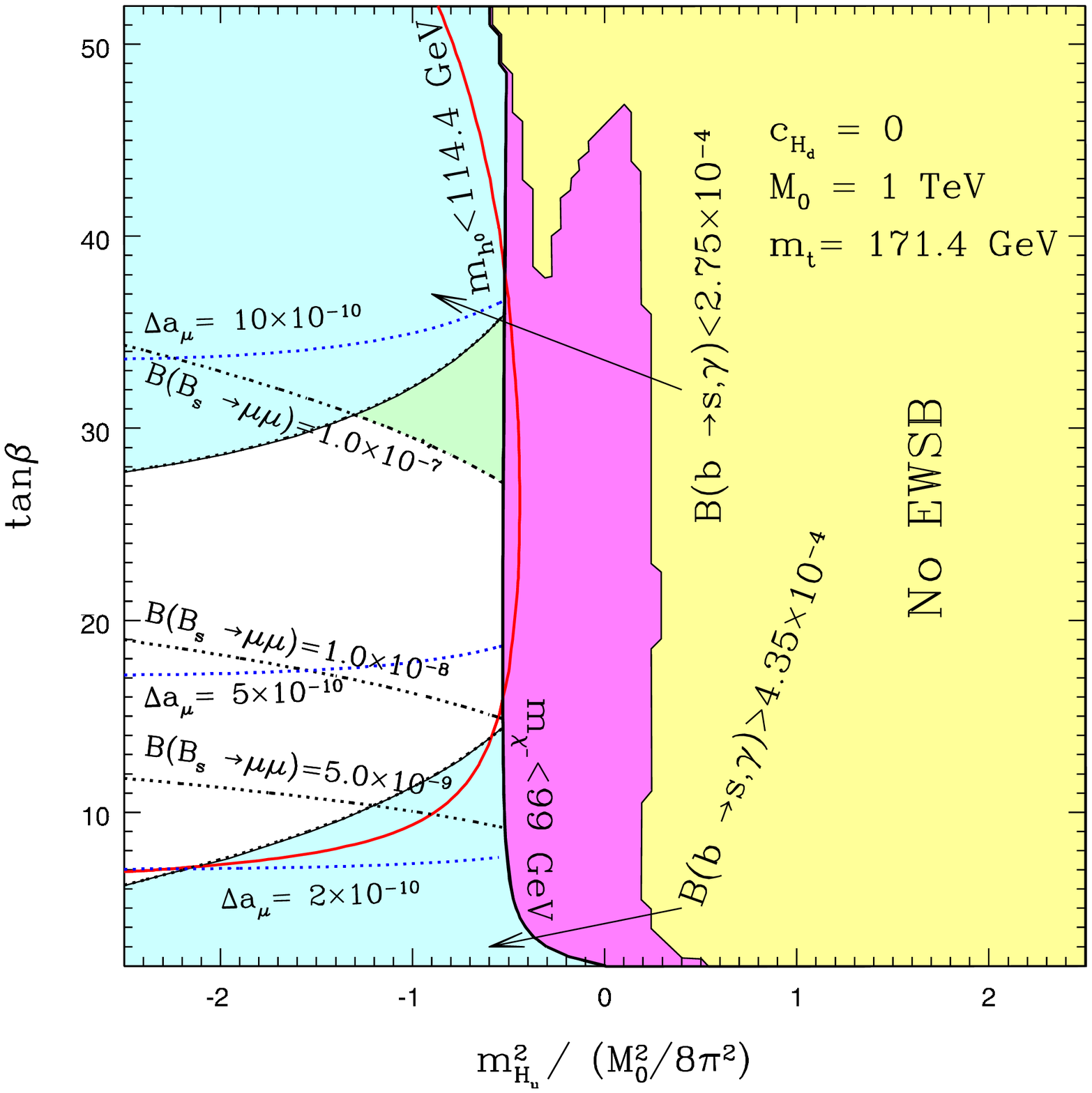,angle=0,width=7.5cm}}
}
\end{minipage}
\caption{Constraints from FCNC and the muon g-2 in the mass
pattern (I). \label{fig:fcnc1}}
\end{center}
\end{figure}

\begin{figure}[t]
\begin{center}
\begin{minipage}{15cm}
\centerline{
{\hspace*{-.2cm}\psfig{figure=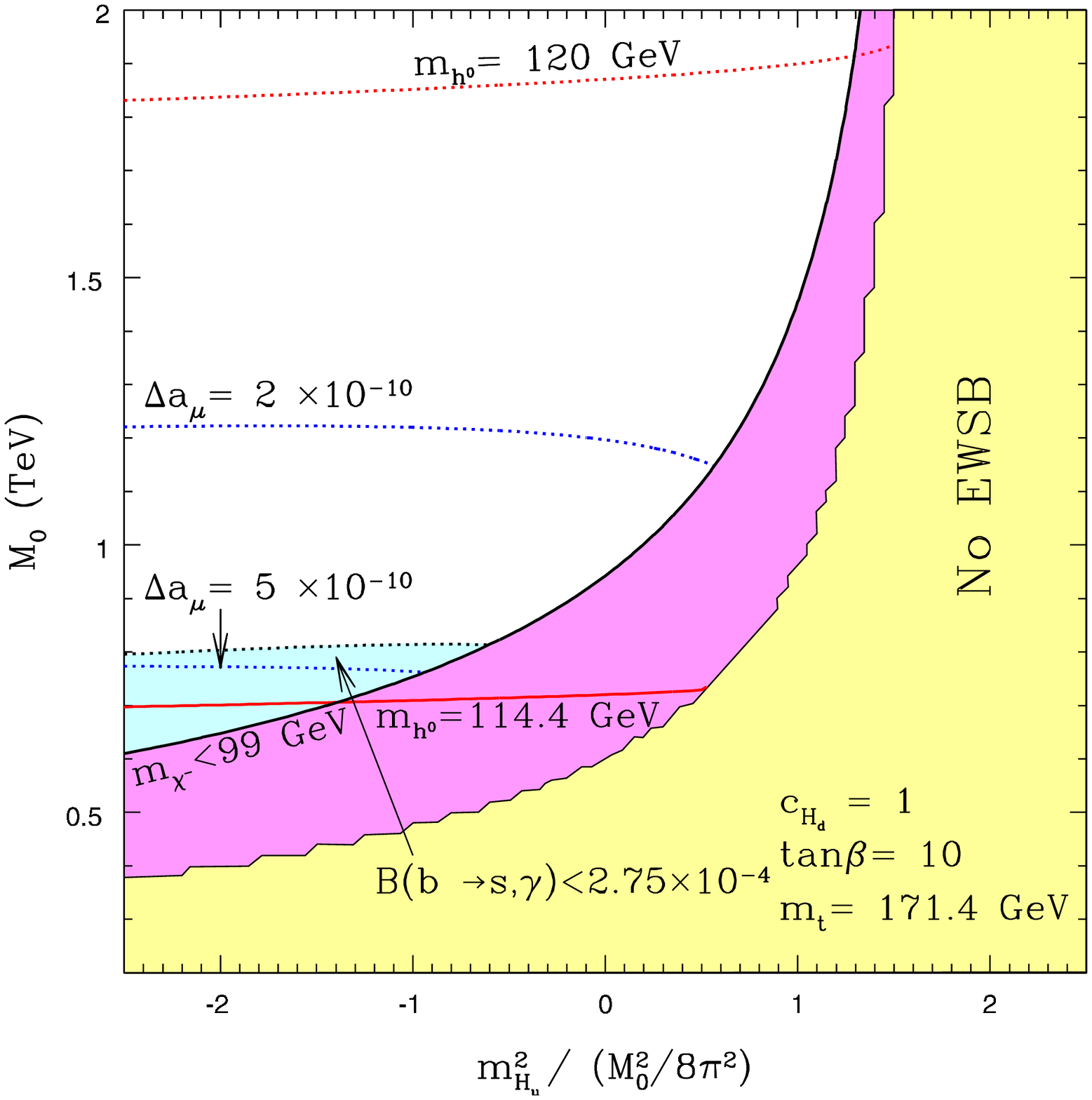,angle=0,width=7.5cm}}
{\hspace*{-.2cm}\psfig{figure=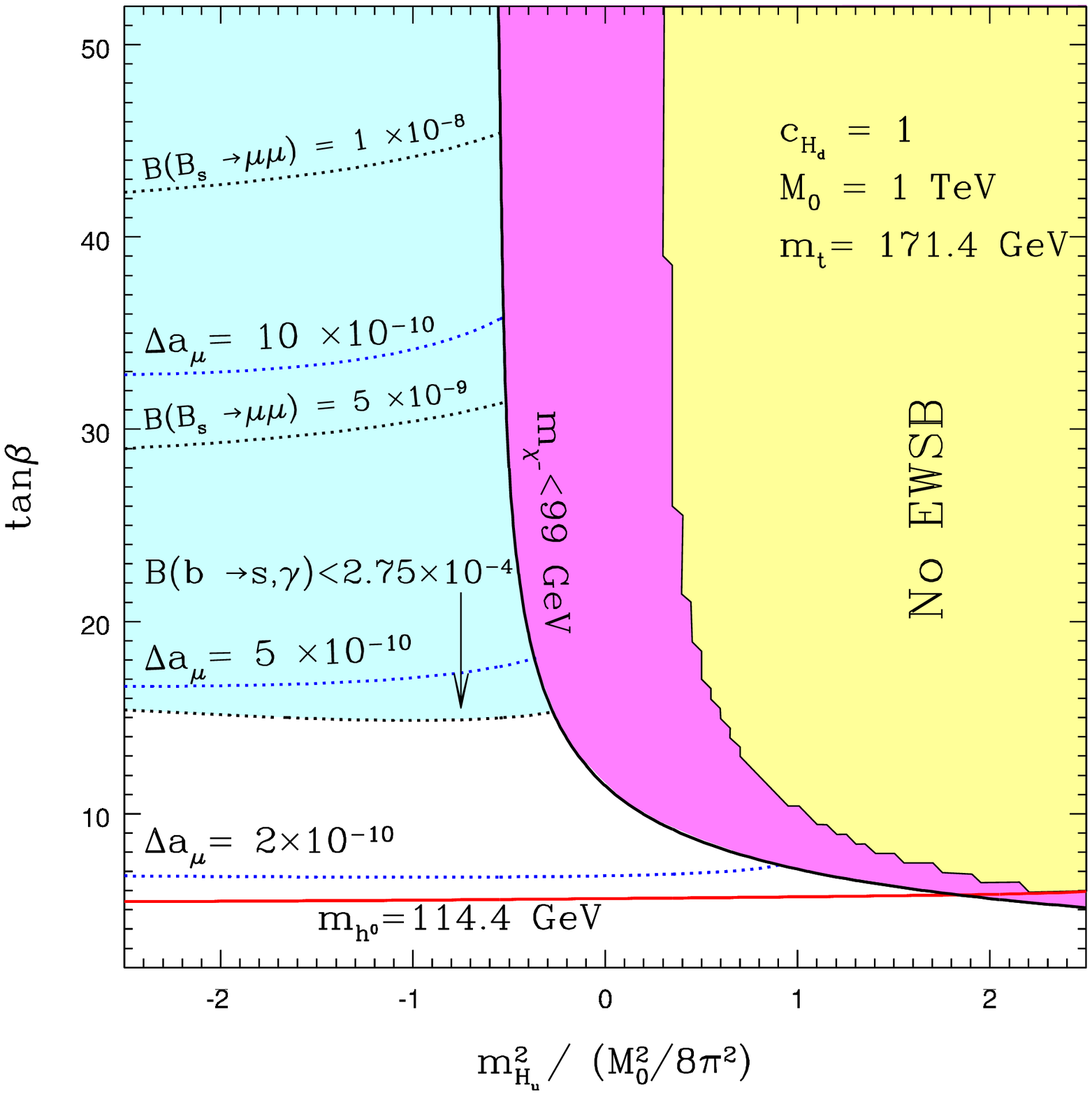,angle=0,width=7.5cm}}
}
\end{minipage}
\caption{Constraints from FCNC and the muon g-2 in the mass
pattern (II). \label{fig:fcnc2}}
\end{center}
\end{figure}

The electroweak symmetry breaking conditions suggest that the TeV scale mirage
mediation solving the little SUSY hierarchy problem can give two different mass
patterns (I) and (II) at the weak scale,
which differ by the values of $m_{H_d}$ and $B$.
In this paper, we analyzed the electroweak symmetry breaking as well as the
constraints from various FCNC processes in both mass patterns.
The results are summarized in Figs.\ref{fig:finetuning1}--\ref{fig:fcnc2},
which show that a large fraction of the parameter space can give the correct
electroweak symmetry breaking while satisfying the FCNC constraints with a reasonable
degree of fine tuning better than 10\%.
For the mass pattern (II), $|\mu|$ can be significantly bigger than $M_Z$ while
keeping the degree of fine tuning better than 10\%,
if $|B|^2\approx m_{H_d}^2$.
This might open up a possibility to raise $|\mu|$, thus raise the Higgsino mass,
without causing a fine tuning problem.

\vskip 1cm

\vspace{5mm} \noindent{\large\bf Acknowledgments}
\vspace{5mm}

K.C. and K.S.J are supported by the KRF Grant funded by the Korean Government
(KRF-2005-201-C00006), the KOSEF Grant (KOSEF R01-2005-000-10404-0),
and the Center for High Energy Physics of Kyungpook National University.
T.K.\/ is supported in part by the Grand-in-Aid for Scientific Research \#17540251,
and the Grant-in-Aid for the 21st Century COE ``The Center for Diversity and
Universality in Physics''
and K.O. is supported by the Grand-in-aid for Scientific Research on Priority Areas
\#441: ``Progress in elementary particle physics of the 21st century through discovery
of Higgs boson and supersymmetry'' \#16081209
from the Ministry of Education, Culture, Sports, Science and Technology of Japan.
K.O. thanks to Yukawa Institute in Kyoto University for use of Altix3700 BX2 by
which much of numerical calculation has been made.

\section*{Appendix A.}

In this appendix, we summarize  the notations and conventions used in this paper.
The quantum effective action in $N=1$ superspace can be written as
\bea
&& \int d^4\theta \left[-3CC^*e^{-K/3}
+ \frac{1}{16}\left(
G_aW^{a\alpha}\frac{D^2}{\partial^2}W^a_\alpha+{\rm h.c.}\right)\right]
+ \left(\,\int d^2\theta\, C^3W+{\rm h.c.}\, \right)
\nonumber \\
&=& \int d^4\theta
\left[\,-3CC^*e^{-K_0/3}+CC^*e^{-K_0/3}Z_i\Phi^{i*}e^{2V^aT_a}\Phi^i
+ \frac{1}{16}\left(\,G_aW^{a\alpha}\frac{D^2}{\partial^2}W^a_\alpha
+ {\rm h.c.}\,\right)\,\right]
\nonumber \\
&& +\,\left(\,\int d^2\theta\,
C^3\left[\,W_0+\frac{1}{6}\lambda_{ijk}\Phi^i\Phi^j\Phi^k\,\right]
+ {\rm h.c.} \,\right)+...,
\eea
where the gauge kinetic terms are written as a $D$-term operator to accommodate the
radiative corrections to gauge couplings, and the ellipsis stands for the irrelevant
higher dimensional operators.
The K\"ahler potential $K$ is expanded as
\bea
K = K_0(T_A,T_A^*)+Z_i(T_A,T_A^*)\Phi^{i*}e^{2V^aT_a}\Phi^i + ...,
\eea
where $V^a$ and $\Phi^i$ denote the visible gauge and matter superfields given by
\bea
\Phi^i &=& \phi^i+\sqrt{2}\,\theta\psi^i+\theta^2F^i,
\nonumber \\
V^a &=& -\theta\sigma^\mu\bar\theta A^a_\mu
-i\bar\theta^2\theta\lambda^a + i\theta^2\bar\theta\bar\lambda^a
+ \frac{1}{2}\theta^2\bar\theta^2 D^a,
\eea
and $T_A=(C,T)$ are the SUSY breaking messengers including the conformal
compensator superfield $C=C_0+\theta^2F^C$ and the modulus superfield
$T=T_0+\sqrt{2}\theta\tilde{T}+\theta^2F^T$.
The radiative corrections due to renormalizable gauge and Yukawa interactions
can be encoded in the matter K\"ahler metric $Z_i$ and the gauge coupling
superfield $G_a$ which is given by
\bea
G_a\,=\,{\rm Re}(f_a)+\Delta G_a,
\eea
where $f_a$ is the holomorphic gauge kinetic function and $\Delta G_a$ includes the
$T_A$-dependent radiative correction to gauge coupling.
The superpotential is expanded as
\bea
W = W_0(T)+\frac{1}{6}\lambda_{ijk}(T)\Phi^i\Phi^j\Phi^k+...,
\eea
where $W_0(T)$ is the modulus superpotential stabilizing $T$.
Here we do not specify the mechanism to generate the MSSM Higgs parameters $\mu$ and
$B$, and treat them as free parameters.

For the canonically normalized component fields, the above superspace action gives
the following form of the running gauge and Yukawa couplings,
the supersymmetric gaugino-matter fermion coupling ${\cal L}_{\lambda\psi}$,
and the soft SUSY breaking terms:
\bea
\frac{1}{g_a^2(Q)} &=&
{\rm Re}(G_a),\quad y_{ijk}(Q) \,=\,
\frac{\lambda_{ijk}}{\sqrt{e^{-K_0}Z_iZ_jZ_k}},
\nonumber \\
{\cal L}_{\lambda\psi} &=&
i\sqrt{2}\left(\phi^{i*} T_a\psi^i\lambda^a
- \bar\lambda^aT_a\phi^i\bar\psi^i \right),
\nonumber \\
{\cal L}_{\rm soft} &=& -m^2_i\phi^i\phi^{i*}
- \left(\,\frac{1}{2}M_a\lambda^a\lambda^a
+ \frac{1}{6}A_{ijk}y_{ijk}\phi^i\phi^j\phi^k + {\rm h.c.}\right),
\eea
where $Q$ denotes the renormalization point and
\bea
M_a(Q) &=& F^A\partial_A\ln ({\rm Re}(G_a)),
\nonumber \\
A_{ijk}(Q) &=& -F^A\partial_A \ln\left(
\frac{\lambda_{ijk}}{e^{-K_0}Z_iZ_jZ_k}\right),
\nonumber \\
m^2_i(Q) &=&  -F^AF^{B*}\partial_A\partial_{\bar B} \ln\left(
e^{-K_0/3}Z_i\right) \eea
for
\bea
F^T &=&
-e^{K_0/2}(\partial_T\partial_{\bar T}K_0)^{-1}(D_TW_0)^*,
\nonumber \\
F^C &=&
m_{3/2}^*+\frac{1}{3}\partial_TK_0F^T \quad(m_{3/2} = e^{K_0/2}W_0).
\eea
In the approximation ignoring the off-diagonal components of
$w_{ij}=\sum_{pq}y_{ipq}y^*_{jpq}$, the 1-loop RG evolution of soft parameters is
determined by
\bea
{16\pi^2}\frac{dM_a}{d\ln Q} &=& 2
\left[-3\,{\rm tr}\Big(T_a^{2}({\rm Adj})\Big)
+ \sum_i {\rm tr}\Big(T_a^{2}(\phi^i)\Big) \right] g^2_aM_a,
\nonumber \\
{16\pi^2}\frac{dA_{ijk}}{d\ln Q} &=& \left[
\sum_{pq}|y_{ipg}|^2A_{ipg} - 4 \sum_a g^2_aC_2^a(\phi^i) M_a
\right] + \Big[i \leftrightarrow j\Big] + \Big[i \leftrightarrow k\Big],
\nonumber \\
{16\pi^2}\frac{d m^2_i}{d\ln Q} &=&
\sum_{jk}|y_{ijk}|^2\left(m^2_i+m^2_j+m^2_k+|A_{ijk}|^2\right)
\nonumber \\
&& -\,
8 \sum_a g^2_aC_2^a(\phi^i)|M_a|^2 +2g_1^2q_i
\sum_j q_j m^2_j,
\eea
where the quadratic Casimir $C^a_2(\phi^i)=(N^2-1)/2N$ for a fundamental
representation $\phi^i$ of the gauge group $SU(N)$, $C_2^a(\phi^i)=q_i^2$ for the
$U(1)$ charge $q_i$ of $\phi^i$.

In mirage mediation, soft terms at $M_{GUT}$ are determined by the modulus mediation
of ${\cal O}\Big(\frac{F^T}{T}\Big)$ and the anomaly mediation of
${\cal O}\Big(\frac{F^C}{8\pi^2 C_0}\Big)$ which are comparable to each other.
In the presence of the axionic shift symmetry
\bea
U(1)_T &:& \quad {\rm Im}(T) \,\rightarrow\, {\rm Im}(T)+ \mbox{real constant}
\eea
which is broken by the non-perturbative term in the modulus superpotential
\bea
W_0 = w-Ae^{-aT},
\eea
one can always make that $m_{3/2}$ and $F^T$ are simultaneously real.
Also since $\frac{F^T}{T}\sim \frac{m_{3/2}}{4\pi^2}$, we have
\bea
\frac{F^C}{C_0} = m_{3/2}\left(\,1
+ {\cal O}\left(\frac{1}{8\pi^2}\right)\,\right).
\eea
Then, upon ignoring the parts of ${\cal O}\Big(\frac{F^T}{8\pi^2 T}\Big)$,
the resulting soft parameters {\it at the scale just below} $M_{GUT}$ are given by
\bea
M_a(M_{GUT}) &=& M_0 + \frac{m_{3/2}}{16\pi^2}\,b_ag_a^2,
\nonumber \\
A_{ijk}(M_{GUT}) &=&
\tilde{A}_{ijk} - \frac{m_{3/2}}{16\pi^2}\,(\gamma_i+\gamma_j+\gamma_k),
\nonumber \\
m_i^2(M_{GUT}) &=&
\tilde{m}_i^2-\frac{m_{3/2}}{16\pi^2}M_0\,\theta_i
- \left(\frac{m_{3/2}}{16\pi^2}\right)^2\dot{\gamma}_i,
\eea
where
\bea
M_0 &=& F^T\partial_T\ln{\rm Re}(f_a),
\nonumber \\
\tilde{A}_{ijk} &=&
-F^T\partial_T\ln\left(\frac{\lambda_{ijk}}{e^{-K_0}Z_iZ_jZ_k}\right)
\,\equiv\, a_{ijk}M_0,
\nonumber \\
\tilde{m}_i^2 &=& -|F^T|^2\partial_T\partial_{\bar{T}}
\ln(e^{-K_0/3}Z_i)\,\equiv\, c_iM_0^2,
\eea
and
\bea
b_a &=&- 3{\rm tr}\left(T_a^2({\rm Adj})\right)
+ \sum_i {\rm tr}\left(T^2_a(\phi^i)\right),
\nonumber \\
\gamma_i &=& 2\sum_a g_a^2C^a_2(\phi^i)-\frac{1}{2}\sum_{jk}|y_{ijk}|^2,
\nonumber \\
\theta_i &=& 4\sum_a g_a^2 C^a_2(\phi^i)-\sum_{jk}a_{ijk}|y_{ijk}|^2,
\nonumber \\
\dot{\gamma}_i &=& 8\pi^2\frac{d\gamma_i}{d\ln Q},
\eea
where $\omega_{ij}=\sum_{kl}y_{ikl}y^*_{jkl}$ is assumed to be diagonal.
Note that if $\lambda_{ijk}$ are $T$-independent constant as required by the axionic
shift symmetry $U(1)_T$, $\tilde{A}_{ijk}= F^T\partial_T\ln(e^{-K_0}Z_iZ_jZ_k)$.

Let us now summarize our conventions for the MSSM.
The superpotential of canonically normalized matter superfields is given by
\bea
W &=& y_DH_d\cdot QD^c+y_LH_d\cdot LE^c-y_UH_u\cdot
QU^c - \mu H_d\cdot H_u,
\eea
where the $SU(2)_L$ product is $H\cdot Q=\epsilon_{ab}H^aQ^b$ with
$\epsilon_{12}=-\epsilon_{21}=1$,
and color indices are suppressed.
Then the chargino and neutralino mass matrices are given by
\bea
-\frac{1}{2}\,\tilde\psi^{-T}{\cal M}_C\tilde\psi^+
-\frac{1}{2}\,\tilde\psi^{0T}{\cal M}_N\tilde\psi^0 + {\rm h.c.},
\eea
where
\bea
{\cal M}_C &=& \left(
\begin{array}{cc}
- M_2\,\, & g_2 \langle H^0_u \rangle \\
g_2 \langle H^0_d \rangle & \mu
\end{array}
\right),
\nonumber \\
{\cal M}_N &=& \left(
\begin{array}{cccc}
-M_1 & 0 & -\frac{1}{\sqrt 2}\,g_Y \langle H^0_d \rangle
& \frac{1}{\sqrt 2}\,g_Y \langle H^0_u \rangle \\
0 & -M_2 & \frac{1}{\sqrt 2}\,g_2 \langle H^0_d \rangle
& -\frac{1}{\sqrt 2}\,g_2 \langle H^0_u \rangle \\
  -\frac{1}{\sqrt 2}\,g_Y \langle H^0_d \rangle
& \frac{1}{\sqrt 2}\,g_2 \langle H^0_d \rangle
& 0 & -\mu  \\
  \frac{1}{\sqrt 2}\,g_Y \langle H^0_u \rangle
& -\frac{1}{\sqrt 2}\,g_2 \langle H^0_u \rangle & -\mu & 0
\end{array}
\right),
\eea
in the field basis
\bea
\tilde\psi^{+T} &=&
-i\left(\tilde W^+,\, i\tilde H^+_u \right), \quad \tilde\psi^{-T}
\,=\, -i\left(\tilde W^-,\, i\tilde H^-_d \right),
\nonumber \\
\tilde\psi^{0T} &=& -i\left( \tilde B,\,\tilde W^3,\, i\tilde
H^0_d,\,i\tilde H^0_u \right),
\eea
for $\tilde W^{\pm}=(\tilde W^1\mp i \tilde W^2)/\sqrt 2.$

The one-loop beta function coefficients $b_a$ and anomalous
dimension $\gamma_i$ in the MSSM are given by \bea b_3 &=& -3,\qquad
b_2\,=\,1,\qquad b_1\,=\,\frac{33}{5},
\nonumber \\
\gamma_{H_u} &=& \frac{3}{2}g_2^2+\frac{1}{2}g_Y^2 -3y_t^2,
\nonumber \\
\gamma_{H_d} &=& \frac{3}{2}g_2^2+\frac{1}{2}g_Y^2 - 3 y_b^2 - y_\tau^2,
\nonumber \\
\gamma_{Q_a} &=& \frac{8}{3} g_3^2 + \frac{3}{2} g_2^2
+ \frac{1}{18} g_Y^2 - (y_t^2 + y_b^2) \delta_{3a},
\nonumber \\
\gamma_{U_a} &=& \frac{8}{3} g_3^2  + \frac{8}{9} g_Y^2 - 2 y_t^2 \delta_{3a},
\nonumber \\
\gamma_{D_a} &=& \frac{8}{3} g_3^2 + \frac{2}{9} g_Y^2 - 2 y_b^2 \delta_{3a},
\nonumber \\
\gamma_{L_a} &=& \frac{3}{2} g_2^2 + \frac{1}{2} g_Y^2 - y_\tau^2 \delta_{3a},
\nonumber \\
\gamma_{E_a} &=& 2 g_Y^2 - 2 y_\tau^2 \delta_{3a},
\eea
where $g_2$ and $g_Y=\sqrt{3/5}\,g_1$ denote the $SU(2)_W$ and $U(1)_Y$ gauge
couplings.
The $\theta_i$ and $\dot{\gamma}_i$ which determine the soft scalar masses at $M_{GUT}$
are given by
\bea
\theta_{H_u} &=&
3g_2^2+g_Y^2 -6y_t^2 a_{H_uQ_3U^c_3},
\nonumber \\
\theta_{H_d} &=& 3g_2^2+g_Y^2 - 6y_b^2 a_{H_dQ_3D^c_3} - 2y_\tau^2 a_{H_dL_3E^c_3},
\nonumber \\
\theta_{Q_a} &=& \frac{16}{3} g_3^2 + 3 g_2^2
+ \frac{1}{9} g_Y^2
- 2\Big(y_t^2a_{H_uQ_3U^c_3} + y_b^2a_{H_dQ_3D^c_3}\Big) \delta_{3a},
\nonumber \\
\theta_{U_a} &=& \frac{16}{3} g_3^2  + \frac{16}{9} g_Y^2
- 4y_t^2a_{H_uQ_3U^c_3} \delta_{3a},
\nonumber \\
\theta_{D_a} &=& \frac{16}{3} g_3^2 + \frac{4}{9} g_Y^2
- 4y_b^2a_{H_dQ_3D^c_3} \delta_{3a},
\nonumber \\
\theta_{L_a} &=& 3 g_2^2 + g_Y^2
- 2y_\tau^2a_{H_dL_3E^c_3} \delta_{3a},
\nonumber \\
\theta_{E_a} &=& 4 g_Y^2 - 4 y_\tau^2 a_{H_dL_3E^c_3} \delta_{3a},
\eea
and
\bea \dot\gamma_{H_u} &=& \frac{3}{2} g_2^4
+ \frac{11}{2} g_Y^4 - 3 y_t^2 b_{y_t},
\nonumber \\
\dot\gamma_{H_d} &=& \frac{3}{2} g_2^4 + \frac{11}{2} g_Y^4
- 3 y_b^2 b_{y_b} - y_\tau^2 b_{y_\tau},
\nonumber \\
\dot \gamma_{Q_a} &=&  -8 g_3^4 + \frac{3}{2} g_2^4 +
\frac{11}{18} g_Y^4
-(y_t^2 b_{y_t} + y_b^2  b_{y_b}) \delta_{3a},
\nonumber \\
\dot\gamma_{U_a} &=& - 8 g_3^4  +  \frac{88}{9} g_Y^4
- 2 y_t^2 b_{y_t} \delta_{3a},
\nonumber \\
\dot\gamma_{D_a} &=& - 8 g_3^4 + \frac{22}{9} g_Y^4
- 2 y_b^2 b_{y_b} \delta_{3a},
\nonumber \\
\dot\gamma_{L_a} &=& \frac{3}{2}g_2^4 + \frac{11}{2} g_Y^4
- y_\tau^2 b_{y_\tau} \delta_{3a},
\nonumber \\
\dot\gamma_{E_a} &=& 22 g_Y^4 - 2 y_\tau^2 b_{y_\tau} \delta_{3a},
\eea where \bea b_{y_t} &=& - \frac{16}{3} g_3^2 - 3 g_2^2 -
\frac{13}{9} g_Y^2 + 6 y_t^2 + y_b^2,
\nonumber \\
b_{y_b} &=& - \frac{16}{3} g_3^2 - 3 g_2^2 - \frac{7}{9} g_Y^2
+ y_t^2 + 6 y_b^2 + y_\tau^2,
\nonumber \\
b_{y_\tau} &=& - 3 g_2^2 - 3 g_Y^2 + 3 y_b^2  + 4 y_\tau^2.
\eea

In our convention of the gaugino masses and $A$-parameters, the 1-loop RG evolution
of the stop trilinear coupling $A_t\equiv A_{H_ut_Lt_R}$ in the MSSM is given by
\bea
\frac{d A_t}{d\ln Q} &=&
\frac{1}{8\pi^2} \left[ 6y^2_tA_t+y^2_bA_b - \left(
\frac{16}{3}g^2_3M_3+3g^2_2M_2+\frac{13}{9}g^2_YM_1\right)
\right].
\eea


\begin{thebibliography}{99}

\bibitem{Nilles:1983ge}
  H.~P.~Nilles,
  Phys.\ Rept.\  {\bf 110}, 1 (1984);
  H.~E.~Haber and G.~L.~Kane,
  Phys.\ Rept.\  {\bf 117}, 75 (1985).

\bibitem{Haber:1990aw}
  H.~E.~Haber and R.~Hempfling,
  Phys.\ Rev.\ Lett.\  {\bf 66}, 1815 (1991);
  Y.~Okada, M.~Yamaguchi and T.~Yanagida,
  Prog.\ Theor.\ Phys.\  {\bf 85}, 1 (1991);
  Phys.\ Lett.\ B {\bf 262}, 54 (1991);
  J.~R.~Ellis, G.~Ridolfi and F.~Zwirner,
  Phys.\ Lett.\ B {\bf 257}, 83 (1991);
  Phys.\ Lett.\ B {\bf 262}, 477 (1991).

\bibitem{Barbieri:1987fn}
  R.~Barbieri and G.~F.~Giudice,
  Nucl.\ Phys.\ B {\bf 306}, 63 (1988);
  P.~H.~Chankowski, J.~R.~Ellis and S.~Pokorski,
  Phys.\ Lett.\ B {\bf 423}, 327 (1998)
  [arXiv:hep-ph/9712234];
  P.~H.~Chankowski, J.~R.~Ellis, M.~Olechowski and S.~Pokorski,
  Nucl.\ Phys.\ B {\bf 544}, 39 (1999)
  [arXiv:hep-ph/9808275];
  G.~L.~Kane and S.~F.~King,
  Phys.\ Lett.\ B {\bf 451}, 113 (1999)
  [arXiv:hep-ph/9810374];
  M.~Bastero-Gil, G.~L.~Kane and S.~F.~King,
  Phys.\ Lett.\ B {\bf 474}, 103 (2000)
  [arXiv:hep-ph/9910506].

\bibitem{casas}
A.~Brignole, J.~A.~Casas, J.~R.~Espinosa and I.~Navarro,
Nucl.\ Phys.\ B {\bf 666}, 105 (2003) [arXiv:hep-ph/0301121];
J.~A.~Casas, J.~R.~Espinosa and I.~Hidalgo,
JHEP {\bf 0401}, 008 (2004) [arXiv:hep-ph/0310137].

\bibitem{Batra:2003nj}
P.~Batra, A.~Delgado, D.~E.~Kaplan and T.~M.~P.~Tait,
JHEP  {\bf 0402}, 043 (2004) [arXiv:hep-ph/0309149].

\bibitem{Harnik:2003rs}
R.~Harnik, G.~D.~Kribs, D.~T.~Larson and H.~Murayama,
Phys.\ Rev.\ D {\bf 70}, 015002 (2004) [arXiv:hep-ph/0311349].

\bibitem{Kobayashi:2004pu}
  T.~Kobayashi and H.~Terao,
  JHEP {\bf 0407}, 026 (2004)
  [arXiv:hep-ph/0403298];
  T.~Kobayashi, H.~Nakano and H.~Terao,
  Phys.\ Rev.\ D {\bf 71}, 115009 (2005)
  [arXiv:hep-ph/0502006];
T.~Kobayashi, H.~Terao and A.~Tsuchiya,
Phys.\ Rev.\ D {\bf 74}, 015002 (2006) [arXiv:hep-ph/0604091].

\bibitem{Chang:2004db}
  S.~Chang, C.~Kilic and R.~Mahbubani,
  Phys.\ Rev.\ D {\bf 71}, 015003 (2005)
  [arXiv:hep-ph/0405267].

\bibitem{Birkedal:2004zx}
A.~Birkedal, Z.~Chacko and Y.~Nomura,
Phys.\ Rev.\ D {\bf 71}, 015006 (2005)
 [arXiv:hep-ph/0408329].

\bibitem{Babu:2004xg}
  K.~S.~Babu, I.~Gogoladze and C.~Kolda,
  arXiv:hep-ph/0410085.

\bibitem{Birkedal:2004xi}
  A.~Birkedal, Z.~Chacko and M.~K.~Gaillard,
  JHEP {\bf 0410}, 036 (2004)
  [arXiv:hep-ph/0404197].

\bibitem{radovan}
R. Dermisek and J. F. Gunion, Phys. Rev. Lett. {\bf 95}, 041801
(2005) [arXiv:hep-ph/0502105]; R. Dermisek and J. F. Gunion, Phys.
Rev. D{\bf 73}, 111701 (2006) [arXiv:hep-ph/0510322].

\bibitem{Chacko:2005ra}
  Z.~Chacko, Y.~Nomura and D.~Tucker-Smith,
  Nucl.\ Phys.\ B {\bf 725}, 207 (2005)
  [arXiv:hep-ph/0504095].

\bibitem{Delgado:2005fq}
  A.~Delgado and T.~M.~P.~Tait,
  JHEP {\bf 0507}, 023 (2005)
  [arXiv:hep-ph/0504224].

\bibitem{Kim:2006mb}
  S.~G.~Kim, N.~Maekawa, A.~Matsuzaki, K.~Sakurai, A.~I.~Sanda and T.~Yoshikawa,
Phys. Rev. D {\bf 74}, 115016 [arXiv:hep-ph/0609076].

\bibitem{Giudice:2006sn}
  G.~F.~Giudice and R.~Rattazzi,
  Nucl.\ Phys.\ B {\bf 757}, 19 (2006)
  [arXiv:hep-ph/0606105].

\bibitem{twin-Higgs}
  A.~Falkowski, S.~Pokorski and M.~Schmaltz,
  Phys.\ Rev.\ D {\bf 74}, 035003 (2006)
  [arXiv:hep-ph/0604066];
  S.~Chang, L.~J.~Hall and N.~Weiner, Phys. Rev. D {\bf 75}, 035009
  (2007)
  [arXiv:hep-ph/0604076].

\bibitem{kaplan}
L. M. Carpenter, D. E. Kaplan, E.-J. Rhee, arXiv:hep-ph/0607204.

\bibitem{Choi:2005uz}
  K.~Choi, K.~S.~Jeong and K.~i.~Okumura,
  JHEP {\bf 0509}, 039 (2005)
  [arXiv:hep-ph/0504037].

\bibitem{hdkim}
R. Dermisek and H. D. Kim, Phys.\ Rev.\ Lett. {\bf 96}, 211803
(2006) [arXiv:hep-ph/0601036].

\bibitem{Choi:2005hd}
  K.~Choi, K.~S.~Jeong, T.~Kobayashi and K.~i.~Okumura,
  Phys. Lett. B {\bf 633}, 355 (2006)
  [arXiv:hep-ph/0508029].

\bibitem{Kitano:2005wc}
  R.~Kitano and Y.~Nomura,
  Phys.\ Lett.\ B {\bf 631}, 58 (2005)
  [arXiv:hep-ph/0509039].

\bibitem{choi1}
  K.~Choi, A.~Falkowski, H.~P.~Nilles, M.~Olechowski and S.~Pokorski,
  JHEP {\bf 0411}, 076 (2004)
  [arXiv:hep-th/0411066];
  K.~Choi, A.~Falkowski, H.~P.~Nilles and M.~Olechowski,
  Nucl. Phys. {\bf B718}, 113 (2005) [arXiv:hep-th/0503216].

\bibitem{anomaly}
  L.~Randall and R.~Sundrum,
  Nucl.\ Phys.\ B {\bf 557}, 79 (1999)
  [arXiv: hep-th/9810155];
  G.~F.~Giudice, M.~A.~Luty, H.~Murayama and R.~Rattazzi,
  JHEP {\bf 9812}, 027 (1998)
  [arXiv: hep-ph/9810442];
J. A. Bagger, T. Moroi and E. Poppitz, JHEP {\bf 0004}, 009 (2000)
[arXiv: hep-th/9911029];
  P.~Binetruy, M.~K.~Gaillard and B.~D.~Nelson,
  Nucl.\ Phys.\ B {\bf 604}, 32 (2001)
  [arXiv: hep-ph/0011081].

\bibitem{modulus}
V.~S.~Kaplunovsky and J.~Louis,
Phys.\ Lett.\ B {\bf 306}, 269 (1993) [arXiv: hep-th/9303040];
A.~Brignole, L.~E.~Ibanez and C.~Munoz,
Nucl.\ Phys.\ B {\bf 422}, 125 (1994) [Erratum-ibid.\ B {\bf 436},
747 (1995)] [arXiv: hep-ph/9308271].

\bibitem{hdkim1} R. Dermisek,  H. D. Kim and I.-W. Kim, JHEP {\bf 0610}, 001
(2006) [arXiv:hep-ph/0607169].

\bibitem{nomura1}
  Y. Nomura and D. Poland, hep-ph/0611249.

\bibitem{Kachru:2003aw}
S.~Kachru, R.~Kallosh, A.~Linde and S.~P.~Trivedi,
Phys.\ Rev.\ D {\bf 68}, 046005 (2003) [arXiv:hep-th/0301240].

\bibitem{Kitano:2005ew}
  R.~Kitano and Y.~Nomura,
  B {\bf 632}, 162 (2006)
  [arXiv:hep-ph/0509221];
  R.~Kitano and Y.~Nomura,
  Phys.\ Rev.\ D {\bf 73}, 095004 (2006)
  [arXiv:hep-ph/0602096].

\bibitem{pierce}
  A.~Pierce and J.~Thaler, JHEP {\bf 09}, 017 (2006) [arXiv:hep-ph/0604192].

\bibitem{Abe:2005rx}
  H.~Abe, T.~Higaki and T.~Kobayashi,
  D {\bf 73}, 046005 (2006)
  [arXiv:hep-th/0511160];
  K.~Choi and K.~S.~Jeong, JHEP {\bf 08}, 007 (2006)
  [arXiv:hep-th/0605108];
  K.~Choi, K.~Y.~Lee, Y.~Shimizu, Y.~G.~Kim and K.~i.~Okumura,
  JCAP {\bf 12}, 017 (2006)
  [arXiv:hep-ph/0609132].

\bibitem{susycp}
  K.~Choi,
  Phys.\ Rev.\ Lett.\  {\bf 72}, 1592 (1994)
  [arXiv:hep-ph/9311352].




\bibitem{hebecker}
  F.~Brummer, A.~Hebecker and M.~Trapletti,
  Nucl.\ Phys.\ B {\bf 755}, 186 (2006)
  [arXiv:hep-th/0605232].

\bibitem{fuplifting}
M. Gomez-Reino and C. A. Scrucca, JHEP {\bf 05}, 015 (2006)
[arXiv:hep-th/0602246]; O.~Lebedev, H.~P.~Nilles and M.~Ratz,
  Phys.\ Lett.\ B {\bf 636} (2006) 126
  [arXiv:hep-th/0603047];
 E.~Dudas, C.~Papineau and S.~Pokorski,
  arXiv:hep-th/0610297;
H.~Abe, T.~Higaki, T.~Kobayashi and Y.~Omura,
  Phys.\ Rev.\  D {\bf 75} (2007) 025019
  [arXiv:hep-th/0611024];
R.~Kallosh and A.~Linde, JHEP {\bf 0702}, 002 (2007)
  [arXiv:hep-th/0611183].



\bibitem{hebecker1}
F. Brummer, A. Hebecker and E. Trincherini Nucl. Phys. {\bf B738},
283 (2006) [arXiv:hep-th/0510113].



\bibitem{giddings}
S. B. Giddings and A. Maharana, Phys. Rev. D {\bf 73}, 126003 (2006)
[arXiv:hep-th/0507158].

\bibitem{kachru_sundrum}
S. Kachru, L. McAllister and R. Sundrum, hep-th/0703105.


\bibitem{dine}
A. Anisimov, M. Dine, M. Graesser and S. D. Thomas, Phys. Rev. {\bf
D65}, 105011 (2002) [arXiv:hep-th/0111235]; JHEP {\bf 0203}, 036
(2002) [arXiv:hep-th/0201256].


\bibitem{falkowski}
A. Falkowski, H. M. Lee and C. Ludeling, JHEP {\bf 10}, 090 (2005)
[arXiv:hep-th/0504091].


\bibitem{conformalsequestering}
T. Kobayashi and H. Terao, Phys. Rev. {\bf D64}, 075003 (2001)
[arXiv:hep-ph/0103028]; M. A. Luty and R. Sundrum, Phy. Rev. {\bf
D65}, 066004 (2002) [arXiv:hep-th/0105137];
M. Ibe, K. I. Izawa, Y. Nakayama, Y. Shinbara and T. Yanagida, Phys.
Rev. {\bf D73}, 015004 (2006) [arXiv:hep-ph/0506203];
 M. Schmaltz and R. Sundrum, JHEP {\bf 0611}, 011 (2006) [arXiv:hep-th/0608051].



\bibitem{luty_sundrum}
M. A. Luty and R. Sundrum, Phys. Rev. {\bf D64}, 065012 (2001)
[arXiv:hep-th/0012158].


\bibitem{choi3}
K.~Choi and K.~S.~Jeong, JHEP {\bf 08}, 007 (2006)
  [arXiv:hep-th/0605108].

\bibitem{ks}
I. R. Klebanov and M. J. Strassler, JHEP {\bf 08}, 052 (2000)
[arXiv:hep-th/0007191].


\bibitem{ckn}
  K.~Choi, J.~E.~Kim and H.~P.~Nilles,
  Phys.\ Rev.\ Lett.\  {\bf 73}, 1758 (1994)
  [arXiv:hep-ph/9404311];
%
  K.~Choi, J.~S.~Lee and C.~Munoz,
  Phys.\ Rev.\ Lett.\  {\bf 80}, 3686 (1998)
  [arXiv:hep-ph/9709250];
%
  S.~P.~de Alwis,
  arXiv:hep-th/0607148.

\bibitem{choi_uni}
K. Choi, Phys. Lett. B {\bf 642}, 404 (2006) [arXiV:hep-th/0606104].

\bibitem{ck}
  K.~Choi and J.~E.~Kim,
  Phys.\ Lett.\ B {\bf 165}, 71 (1985);
%
  L.~E.~Ibanez and H.~P.~Nilles,
  Phys.\ Lett.\ B {\bf 169}, 354 (1986);
%
  E.~Witten,
  Nucl.\ Phys.\ B {\bf 471}, 135 (1996)
  [arXiv:hep-th/9602070];
%
  T.~Banks and M.~Dine,
  Nucl.\ Phys.\ B {\bf 479}, 173 (1996)
  [arXiv:hep-th/9605136];
%
  K.~Choi,
  Phys.\ Rev.\ D {\bf 56}, 6588 (1997)
  [arXiv:hep-th/9706171].

\bibitem{lust}
D.~Lust, P.~Mayr, R.~Richter and S.~Stieberger, Nucl.\ Phys. {\bf
B696}, 205 (2004) [arXiv:hep-th/0404134]; M.~Bertolini et. al.,
Nucl.\ Phys. {\bf B743}, 1 (2006) [arXiv:hep-th/0512067].

\bibitem{GKP}
S.~B.~Giddings, S.~Kachru and J.~Polchinski, Phys. Rev. D {\bf 66},
106006 (2002) [arXiv:hep-th/0105097].



\bibitem{ads}
I. Affleck, M. Dine and N. Seiberg, Phys. Rev. Lett. {\bf 51}, 1026
(1983).

\bibitem{adam}
A. Falkowski, O. Lebedev and Y. Mambrini, JHEP {\bf 0511}, 034
(2005) [arXiv:hep-ph/0507110]; O. Lebedev, H. P. Nilles, M. Ratz,
hep-ph/0511320.

\bibitem{munoz}
J. A. Casas, A. Lleyda and C. Munoz, Nucl. Phys. {\bf B471}, 3
(1996) [arXiv:hep-ph/9507294].

\bibitem{kuzenko}
A. Kusenko, P. Langacker and G. Segre, Phys. Rev. {\bf D54}, 5824
(1996) [arXiv:hep-ph/9602414]; A. Kusenko and P. Langacker, Phys.
Lett. B {\bf 391}, 29 (1997) [arXiv:hep-ph/9608340].

\bibitem{falk}
T. Falk, K. A. Olive, L. Roszkowski, A. Singh and M. Srednicki,
Phys. Lett. B {\bf 396}, 50 (1997) [arXiv:hep-ph/9611325].


\bibitem{riotto}
A. Riotto and E. Roulet, Phys. Lett. B {\bf 377}, 60 (1996)
[arXiv:hep-ph/9512401].

\bibitem{Coleman:1973jx}
  S.~R.~Coleman and E.~Weinberg,
  Phys.\ Rev.\ D {\bf 7}, 1888 (1973);
  S.~Weinberg,
  Phys.\ Rev.\ D {\bf 7}, 2887 (1973).

\bibitem{Gamberini:1989jw}
  G.~Gamberini, G.~Ridolfi and F.~Zwirner,
  Nucl.\ Phys.\ B {\bf 331}, 331 (1990);
  B.~de Carlos and J.~A.~Casas,
  Phys.\ Lett.\ B {\bf 309}, 320 (1993)
  [arXiv:hep-ph/9303291].

\bibitem{Barger:1993gh}
  V.~D.~Barger, M.~S.~Berger and P.~Ohmann,
  Phys.\ Rev.\ D {\bf 49}, 4908 (1994)
  [arXiv:hep-ph/9311269];
  D.~M.~Pierce, J.~A.~Bagger, K.~T.~Matchev and R.~j.~Zhang,
  Nucl.\ Phys.\ B {\bf 491}, 3 (1997)
  [arXiv:hep-ph/9606211].

\bibitem{Siegel:1979wq}
  W.~Siegel,
  Phys.\ Lett.\ B {\bf 84}, 193 (1979);
  D.~M.~Capper, D.~R.~T.~Jones and P.~van Nieuwenhuizen,
  Nucl.\ Phys.\ B {\bf 167}, 479 (1980).

\bibitem{FeynHiggs}
  S.~Heinemeyer, W.~Hollik and G.~Weiglein,
  Comput.\ Phys.\ Commun.\  {\bf 124}, 76 (2000)
  [arXiv:hep-ph/9812320].

\bibitem{unknown:2006bi}
  E.~Barberio {\it et al.}  [Heavy Flavor Averaging Group (HFAG)],
  arXiv:hep-ex/0603003.

\bibitem{Gambino:2001ew}
  P.~Gambino and M.~Misiak,
  Nucl.\ Phys.\ B {\bf 611}, 338 (2001)
  [arXiv:hep-ph/0104034];
  A.~J.~Buras, A.~Czarnecki, M.~Misiak and J.~Urban,
  Nucl.\ Phys.\ B {\bf 631}, 219 (2002)
  [arXiv:hep-ph/0203135].

\bibitem{Neubert:2004dd}
  M.~Neubert,
  Eur.\ Phys.\ J.\ C {\bf 40}, 165 (2005)
  [arXiv:hep-ph/0408179].

\bibitem{Becher:2006pu}
  T.~Becher and M.~Neubert,
  Phys.\ Rev.\ Lett.\  {\bf 98}, 022003 (2007)
  [arXiv:hep-ph/0610067].

\bibitem{Huang:1998vb}
  C.~S.~Huang, W.~Liao and Q.~S.~Yan,
  Phys.\ Rev.\ D {\bf 59}, 011701 (1999)
  [arXiv:hep-ph/9803460];
  C.~Hamzaoui, M.~Pospelov and M.~Toharia,
  Phys.\ Rev.\ D {\bf 59}, 095005 (1999)
  [arXiv:hep-ph/9807350];
  S.~R.~Choudhury and N.~Gaur,
  Phys.\ Lett.\ B {\bf 451}, 86 (1999)
  [arXiv:hep-ph/9810307].

\bibitem{Babu:1999hn}
  K.~S.~Babu and C.~F.~Kolda,
  Phys.\ Rev.\ Lett.\  {\bf 84}, 228 (2000)
  [arXiv:hep-ph/9909476];
  C.~S.~Huang, W.~Liao, Q.~S.~Yan and S.~H.~Zhu,
  Phys.\ Rev.\ D {\bf 63}, 114021 (2001)
  [Erratum-ibid.\ D {\bf 64}, 059902 (2001)]
  [arXiv:hep-ph/0006250];
  P.~H.~Chankowski and L.~Slawianowska,
  Phys.\ Rev.\ D {\bf 63}, 054012 (2001)
  [arXiv:hep-ph/0008046];
  C.~Bobeth, T.~Ewerth, F.~Kruger and J.~Urban,
  Phys.\ Rev.\ D {\bf 64}, 074014 (2001)
  [arXiv:hep-ph/0104284];
  Phys.\ Rev.\ D {\bf 66}, 074021 (2002)
  [arXiv:hep-ph/0204225];
  G.~Isidori and A.~Retico,
  JHEP {\bf 0209}, 063 (2002)
  [arXiv:hep-ph/0208159].

\bibitem{Buras:2001mb}
  A.~J.~Buras, P.~H.~Chankowski, J.~Rosiek and L.~Slawianowska,
  Nucl.\ Phys.\ B {\bf 619}, 434 (2001)
  [arXiv:hep-ph/0107048];
%
  Phys.\ Lett.\ B {\bf 546}, 96 (2002)
  [arXiv:hep-ph/0207241];
  Nucl.\ Phys.\ B {\bf 659}, 3 (2003)
  [arXiv:hep-ph/0210145];
  G.~Isidori and A.~Retico,
  JHEP {\bf 0111}, 001 (2001)
  [arXiv:hep-ph/0110121];
%
  J.~Foster, K.~Okumura and L.~Roszkowski,
  Phys.\ Lett.\ B {\bf 609}, 102 (2005)
  [arXiv:hep-ph/0410323];
 JHEP {\bf 0508}, 094 (2005)
 [arXiv:hep-ph/0506146];
  JHEP {\bf 0603}, 044 (2006)
  [arXiv:hep-ph/0510422].

\bibitem{Abazov:2006dm}
  V.~M.~Abazov {\it et al.}  [D0 Collaboration],
  [arXiv:hep-ex/0603029];
  A.~Abulencia  [CDF - Run II Collaboration],
  arXiv:hep-ex/0606027.

\bibitem{Ball:2006xx}
  P.~Ball and R.~Fleischer,
  Eur.\ Phys.\ J.\ C {\bf 48}, 413 (2006)
  [arXiv:hep-ph/0604249].

\bibitem{Carena:2006ai}
  M.~Carena, A.~Menon, R.~Noriega-Papaqui, A.~Szynkman and C.~E.~M.~Wagner,
  Phys.\ Rev.\ D {\bf 74}, 015009 (2006)
  [arXiv:hep-ph/0603106].
\bibitem{Foster:2006ze}
J.~Foster, K.~i.~Okumura and L.~Roszkowski,
Phys.\ Lett.\ B {\bf 641}, 452 (2006) [arXiv:hep-ph/0604121].

\bibitem{Davier:2003pw}
  M.~Davier, S.~Eidelman, A.~Hocker and Z.~Zhang,
  Eur.\ Phys.\ J.\ C {\bf 31}, 503 (2003)
  [arXiv:hep-ph/0308213];
  F.~Jegerlehner,
  Nucl.\ Phys.\ Proc.\ Suppl.\  {\bf 126}, 325 (2004)
  [arXiv:hep-ph/0310234];
  Nucl.\ Phys.\ Proc.\ Suppl.\  {\bf 131}, 213 (2004)
  [arXiv:hep-ph/0312372];
  K.~Hagiwara, A.~D.~Martin, D.~Nomura and T.~Teubner,
  Phys.\ Rev.\ D {\bf 69}, 093003 (2004)
  [arXiv:hep-ph/0312250];
  K.~Melnikov and A.~Vainshtein,
  Phys.\ Rev.\ D {\bf 70}, 113006 (2004)
  [arXiv:hep-ph/0312226];
  J.~F.~de Troconiz and F.~J.~Yndurain,
  Phys.\ Rev.\ D {\bf 71}, 073008 (2005)
  [arXiv:hep-ph/0402285];
  A.~Hocker,
  arXiv:hep-ph/0410081.

\bibitem{Passera:2004bj}
  M.~Passera,
  J.\ Phys.\ G {\bf 31}, R75 (2005)
  [arXiv:hep-ph/0411168].

\bibitem{Bennett:2004pv}
  G.~W.~Bennett {\it et al.}  [Muon g-2 Collaboration],
  Phys.\ Rev.\ Lett.\  {\bf 92}, 161802 (2004)
  [arXiv:hep-ex/0401008].

\bibitem{Hagiwara:2006jt}
  K.~Hagiwara, A.~D.~Martin, D.~Nomura and T.~Teubner,
  arXiv:hep-ph/0611102.

\bibitem{choi4}
K. Choi, K. Hwang, S. K. Kang, K. Y. Lee and W. Y. Song, Phys. Rev.
D{\bf 64}, 055001 (2001) [arXiv:hep-ph/0103048].

\end{thebibliography}
\end{document}